\PassOptionsToClass{
reprint,
superscriptaddress,
nofootinbib,
amsmath, amssymb,
aps,
prl,
floatfix
}{revtex4-2}
\documentclass{revtex4-2}

\usepackage{graphicx}
\usepackage{dcolumn}
\usepackage{bm}
\usepackage{enumitem} 
\usepackage[colorlinks=true, allcolors=blue]{hyperref}
\usepackage[normalem]{ulem} 

\usepackage{bibunits}
\defaultbibliography{bibliography}
\defaultbibliographystyle{apsrev4-2}

\usepackage{xspace}

\newcommand{\lsection}[1]{\textbf{#1.}}

\graphicspath{{figures/}}
\usepackage[separate-uncertainty=false, print-unity-mantissa = false]{siunitx}
\usepackage{braket}

\usepackage[capitalise]{cleveref}

\usepackage{xcolor}

\begin{document}

\title{Dynamical stabilisation of a quantum fluid using a single topological defect}

\date{\today}

\author{Deborah Capecchi}
\author{Paolo Comaron}
\author{Antonio Gianfrate}
\author{Milena De Giorgi}
\author{Dario Ballarini}
\author{Daniele Sanvitto}
\affiliation{CNR Nanotec, Institute of Nanotechnology, via Monteroni, 73100, Lecce, Italy}

\author{Franco Dalfovo}
\affiliation{Pitaevskii BEC Center, CNR-INO and Dipartimento di Fisica, Universit\`a di Trento, 38123 Trento, Italy }

\author{Dimitrios Trypogeorgos}
\email[]{dimitrios.trypogeorgos@cnr.it}
\affiliation{CNR Nanotec, Institute of Nanotechnology, via Monteroni, 73100, Lecce, Italy}



\begin{abstract}
The size and shape of a quantum fluid in equilibrium is strongly influenced by the inter-particle interactions.
In the attractive interaction regime, atomic quantum fluids ultimately collapse in a violent process that expels most of the particles from the macroscopically occupied state.
Here, we use a quantum fluid of light in propagating geometry as an analogue to a two-dimensional Bose-Einstein condensate (BEC) with large attractive interactions to show that non-trivial topology significantly alters the dynamical behaviour of the collapse, enhancing the BEC stability and delaying the collapse time by an order of magnitude.
We measure direct experimental signatures of topology affecting quantum hydrodynamics, unveiling the inherent competition between attractive nonlinearities, that lead to the collapse, and the preservation of topological charge from a multi-charged vortex.
We fully characterise the collapse process in coordinate space and the eventual `solitonification' of the system and connect it to the mode structure of the excitation spectrum.

\end{abstract}

\maketitle


\begin{bibunit}
Topological defects are ubiquitous in nature, manifesting themselves in various contexts such as condensed matter~\cite{Mermin1979,nelson2002}, cosmology~\cite{Kibble1976}, and even in biological systems~\cite{Ardaseva2022, Shankar2022}.  
Their presence can be of particular interest in intrinsically unstable or metastable systems, as the conservation of the topological charge, which makes the defect long-lived, can compete with the system's collapse.
Such a situation is realized by quantized vortices in Bose-Einstein condensates (BECs) with attractive interaction. 
The irrotational nature of the BEC leads to the discretization and conservation of the vortex charge that can be treated as a topological invariant~\cite{Jeffrey2017}. 
In the absence of defects, the attractive interaction induces the collapse of the condensate with spectacular dynamics both in bosonic~\cite{Sackett1998,Roberts2001,Cornish2006,Banerjee2024}, and fermionic~\cite{Modugno2002} systems alike. 
On the contrary, the presence of vortices has been predicted to stabilize the condensate against collapse~\cite{Dalfovo1996, Banerjee2024, Nesti_2026}. 
At the same time, it allows for a competing azimuthal collapse channel~\cite{Bigelow2004, Petrov1998,vuong2006} which complicates the question of the overall BEC stability.

\begin{figure*}
    \centering    
    \includegraphics{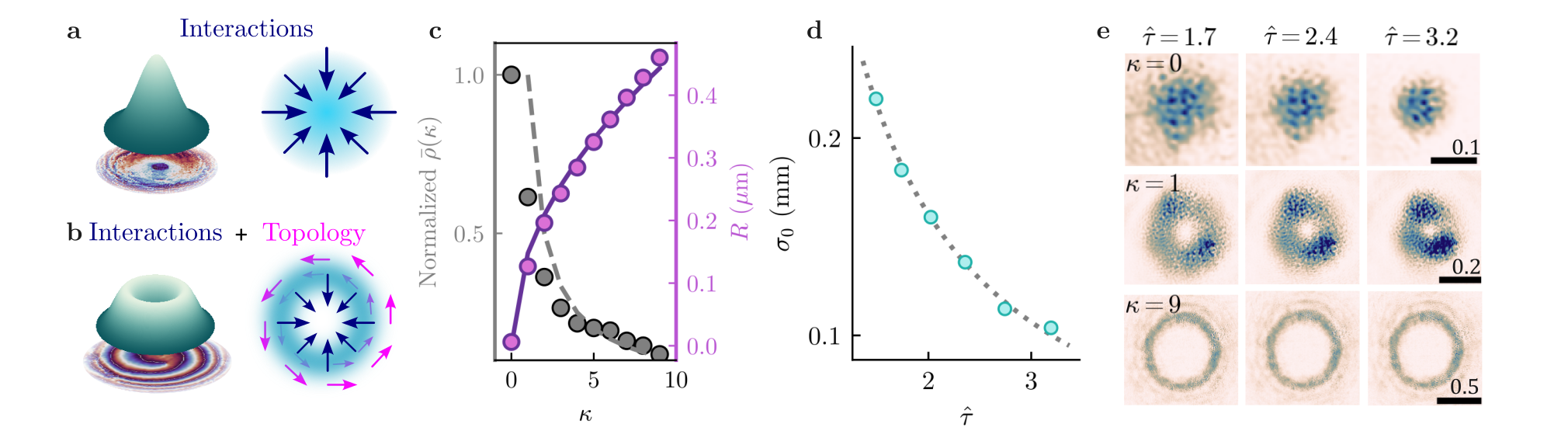}
    \caption{Topological stabilization of the condensate. \textbf{a} For an initially gaussian-shaped condensate, attractive interactions cause a monotonic increase of the density and a corresponding decrease of the width.  \textbf{b} In the presence of a vortex with topological charge $\kappa$, the phase singularity causes the density at the center to vanish, and induces an azimuthal velocity; the measured phase is shown below the density in both \textbf{a} and \textbf{b}.  {\bf c} With increasing vortex charge the measured ring radius $R$ increases and the average ring density $\bar{\rho}(\kappa)$ normalized to the density of the $\kappa=0$ case decreases. Solid and dashed lines correspond to $\sigma_0 \sqrt{\kappa}$ and $1/\kappa$, respectively. \textbf{d} The experimentally measured size $\sigma_0$ of the gaussian-shaped condensate becomes smaller as the $\hat\tau$ increases; the dotted line is a guide to the eye. \textbf{e} The addition of a vortex arrests the decrease as shown in the snapshots for different vortex charges and effective times, the sizes of the scale bars are expressed in mm. }
    \label{fig:concept}
\end{figure*}

Here, we investigate the dynamics of a collapsing two-dimensional quantum fluid of light with multiply charged vortices, in the regime of large attractive interactions, as an analogue of an atomic BEC.
We experimentally and theoretically explore the effects of the competition between the hydrodynamic push of the topological defect and the interaction-induced collapse, and their dependence on the vortex charge and interaction strength. 
Without vortices, the fluid of light collapses, but the process is less violent than in atomic BECs, where inelastic three-body collisions produce heating~\cite{Donley2001}. 
In our case, opposite-sign, higher-order non-linearities limit the increase in density as the system fragments into isolated peaks~\cite{westerberg.PhysRevA.98.053835}.
The addition of a vortex at the center of the fluid stabilizes it against a collapse, and leads to the formation of a ring with a slowly varying radius, as shown in \cref{fig:concept}a,b. 
As interaction increases, a dynamical instability along the ring occurs, eventually resulting in a proliferation of isolated density peaks and finally the fragmentation of the condensate~\cite{2017Nguyen,2003Salasnich,Carr2004} into soliton necklaces. 
This dynamical process can be modeled by the Gross-Pitaevskii theory, thus providing a satisfactory understanding of the entire temporal evolution. 

Our experiment uses a paraxial quantum fluid of light \cite{glorieux2025} realized with a laser beam propagating in a hot atomic vapour in the regime of non-linear light-atom interactions.
Specifically, we address the $5^2\mathrm{S}_{1/2},~F=1 \rightarrow  5^2\mathrm{P}_{3/2}$ of $^{87}$Rb in a spectroscopy cell with natural isotopic abundance, kept at a temperature of $\qty{120}{\celsius}$. 
The Doppler-broadened transition is $\sim$\qty{590}{MHz} wide, while the laser is blue detuned in the range $\lvert\Delta\rvert=3\text{--}1.5\,\mathrm{GHz}$.

In this system, the propagation of the optical field $\mathcal{E}({\bf r},z)$ in the medium can be mapped to the Gross-Pitaevskii equation (GPE) for a two-dimensional condensate~\cite{Carusotto2014,Larre2015} with a macroscopic wavefunction $\psi({\bf r},\tau)$.
The propagation direction $z$ acts as an effective time $\tau$, and the density $\rho(\bf{r},\tau)$ of the field can be identified with the BEC density $|\psi({\bf r},\tau)|^2$.

The nonlinearity parameter $g$ of the GPE is proportional to the third-order Kerr susceptibility of the atoms $\chi^{(3)}=c\epsilon_0n_2$, where $n_2$ is the nonlinear refractive index, and is controlled via the detuning $\Delta$ of the laser from the Doppler-broadened atomic resonance. 
Here $c$ is the speed of light and $\epsilon_0$ the dielectric constant in vacuum. 
We can define an effective adimensional time $\hat\tau = g \rho_0 z \approx 1.4\text{--}6$ with which the wavefunction propagates in the medium, that scales linearly under the effect of the peak density $\rho_0$.

The beam is diffracted on a liquid crystal spatial light modulator (SLM) to control the phase and superimpose a vortex with a topological charge up to $\kappa=9$. 
We image the plane of the SLM at the entrance of the cell, where the nonlinear interactions abruptly change from zero to a negative value; the entrance beam has a waist $\sigma_0\simeq\qty{210}{\mu m}$ and a total power of \qty{\sim 100}{mW}.  
Since the laser detuning is much larger than the broadened transition linewidth, we neglect laser absorption, which would map to single-particle losses in the fluid; for a discussion about the effects of absorption see~\cite{SM}.
During the \qty{7.5}{cm} of propagation in the cell, the phase discontinuity at the center forces the condensate density to vanish and a ring forms. The centrifugal force imparted by the vortex is directly seen in the scaling of the radius. In the linear regime, which is equivalent to optical beam propagation in free space~\cite{Reddy2015}, the spatially-averaged density of the ring $\bar\rho$ decreases with increasing $\kappa$ as $1/\kappa$ and the radius of the ring $R$ increases as $\sigma_0\sqrt{\kappa}$, see \cref{fig:concept}c. 
We interfere the image reconstructed at the cell's output with a reference beam and retrieve the phase and coherent part of the field from the resulting hologram; this allows us to obtain both density and flow information in a single measurement as shown schematically in \cref{fig:concept}a,b. 

\begin{figure}
    \centering 
    \includegraphics{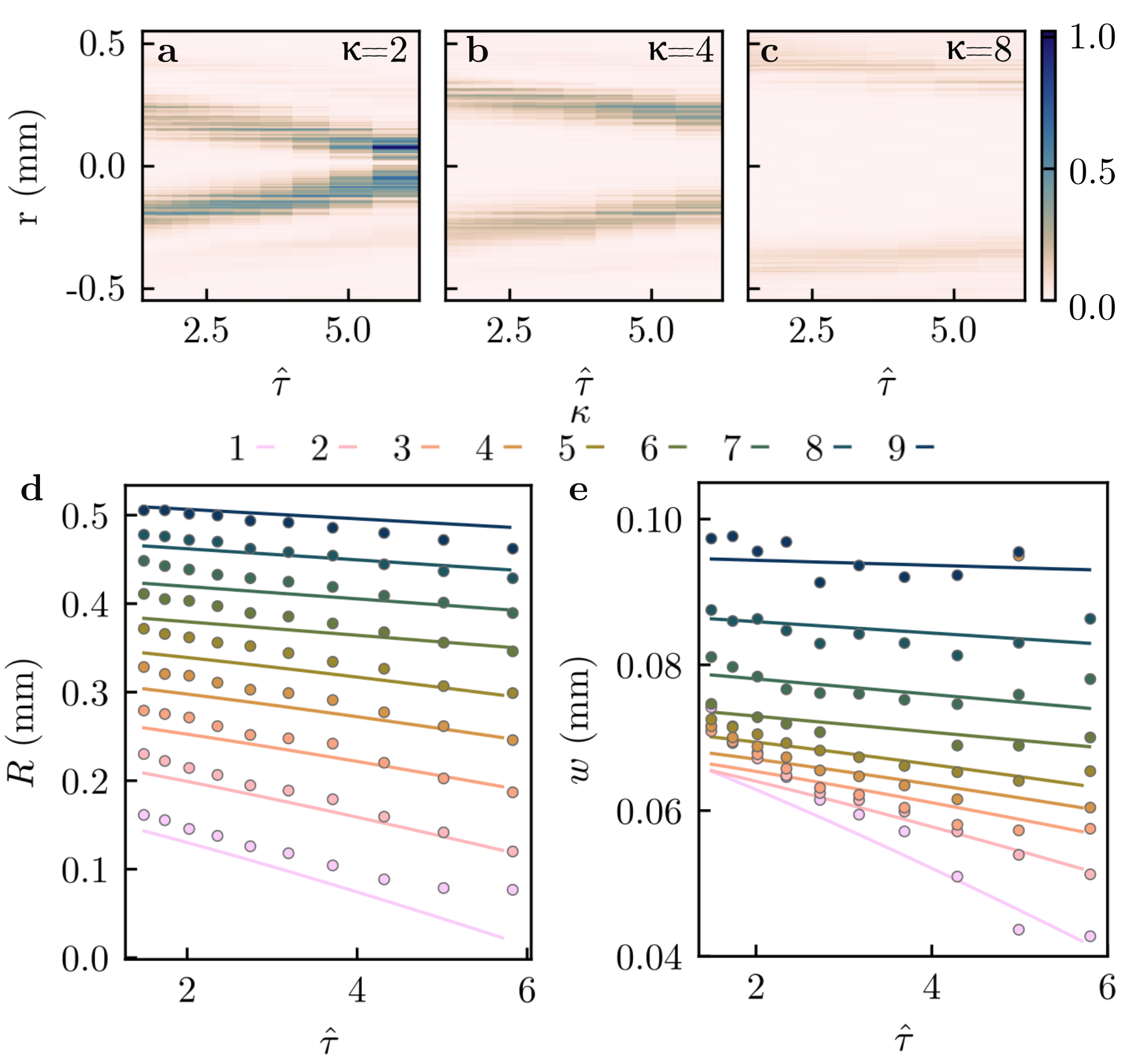}
     \caption{Condensate geometry. \textbf{a-c} Experimental radial profiles of the ring condensate density as a function of $\hat\tau$ for $\kappa=2,4,8$. For large $\kappa$, the evolution for the ring becomes slower and slower, due to a balance between attractive interactions and topological constraints.
    {\bf d}, {\bf e} Radius $R$ and width $w$ of the ring as a function of the effective time $\hat{\tau}$ for different values of  $\kappa$ (legend on top);  circles and lines represent the experimentally measured values and the results of GPE simulations, respectively.
  }
    \label{fig:shape}
\end{figure}

Light propagating in the cell behaves as an interacting BEC with attractive interactions. 
Such a BEC is dynamically unstable because the amplitude of its collective low-energy Bogoliubov excitations can grow unbounded, and eventually fragment the condensate~\cite{Sackett1998}.
The important insight of this work is that, since this process is dynamic, the point of collapse, defined as when the amplitude of the dominant phononic mode becomes comparable to the density of the BEC, can be strongly delayed by engineering the phase of the macroscopic wavefunction.
This becomes apparent by examining the geometric characteristics of the BEC.
The shape of the condensate without the defect is uniquely parametrized by $\sigma_0$. 
The attraction-induced instability manifests as a rapid decrease of $\sigma_0$, shown in \cref{fig:concept}d, and a consequent increase of the central density; this case sets the baseline for the experiments.
The inclusion of even a single quantum vortex drastically changes the topology of the BEC.
Since the value of $\sigma_0$ is always larger than the healing length of the system $\xi\lesssim\qty{8}{\mu m}$, the attractive condensate forms a ring around the phase discontinuity. 
This new topology requires the introduction of two length scales: the radius $R\equiv\braket{R}_\theta$, from the center to the higher-density region of the ring, and the width of the ring $w\equiv\braket{w}_\theta$. 
An azimuthal average accounts for any asymmetries and inhomogeneities of the ring~\cite{SM}.

Even with a single vortex $\kappa=1$, the radial collapse is significantly delayed while it is asymptotically arrested for $\kappa=9$; see \cref{fig:concept}e.
The collapse process can be directly visualized from the radial profiles in \cref{fig:shape}a-c.
The two bands correspond to the two sides of the ring on a plane of intersection. 
For $\kappa = 2$, the ring contracts rapidly under the action of the nonlinearity, keeping the vortex at the center. 
For $\kappa = 4$, the contraction of the ring is significantly slowed down, while for $\kappa = 8$, its radius remains nearly constant.

A systematic analysis of $R$ and $w$ shows that their reduction becomes smaller with increasing $\kappa$, as shown in \cref{fig:shape}d,e.
The ratio of $R$ between the smaller and larger $\hat\tau$ changes from 0.23 for $\kappa=1$ to 0.88 for $\kappa=9$.
The decrease of $w$ can also be understood as the competition between outwards flow imparted by the vortex and the local radial velocity field which tends to point towards the higher density region~\cite{SM}.

\begin{figure}

    \centering
    \includegraphics{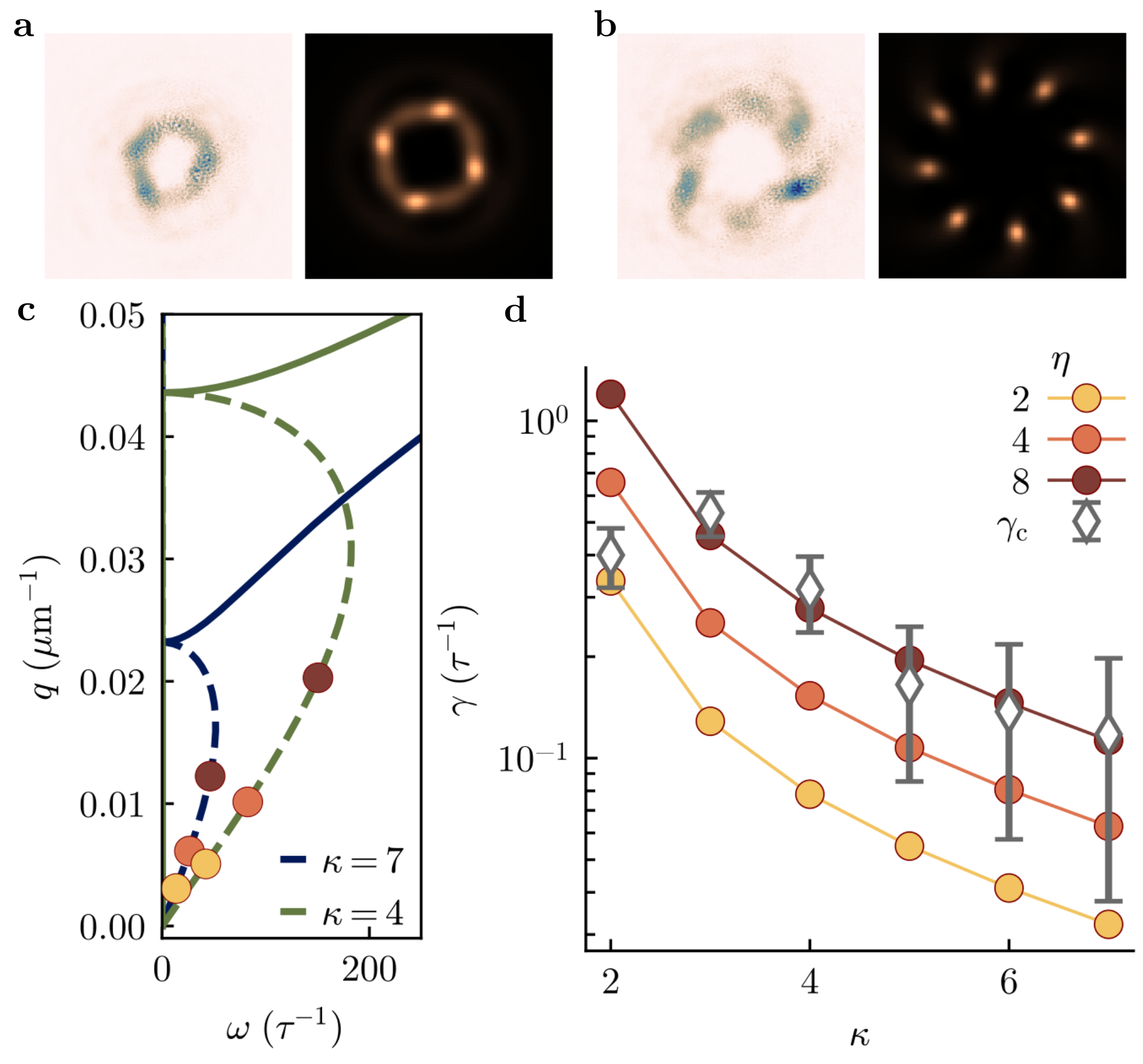}
    \caption{Growth of Bogoliubov modes in the ring. \textbf{a,b} Density profiles from the experiment (left) and numerical simulations with the GPE (right) for $\kappa=4$  at $\hat\tau=8.3$ (a), and $\kappa=7$  at $\hat\tau=13.1$(b). \textbf{c} The frequency of the Bogoliubov modes is purely imaginary at small $q$ and real for large $q$.  The predictions for the corresponding imaginary and real parts are shown as dashed and solid lines, respectively. The modes $\eta$ permitted by the ring periodicity are indicated by colored circles. \textbf{d} Growth rates of the dominant modes (circles) extracted from the Bogoliubov analysis at $\hat\tau=6.8$ are plotted as a function of $\kappa$. They agree well with the growth rate $\gamma_c$ calculated from the GPE simulations (diamonds), except for $\kappa=2$, which has already collapsed in the GPE simulations. The colors correspond to those in panel \textbf{c}.}
        
    \label{fig:collapse}
\end{figure}

For $\kappa \geq 1$, in addition to its radial dynamics, the ring exhibits an azimuthal modulational instability, as shown in \cref{fig:collapse}a,b. 
This can be explained by considering that, when $w \ll R$, the BEC dynamics are driven primarily by Bogoliubov phonons propagating along the ring, on top of slowly varying background.
Due to periodic boundary conditions, azimuthal phonons have discrete wave vectors $q_\eta=\eta/R$, where $\eta$ is an integer.
Counter-propagating phonons with $q_\eta$ and $-q_\eta$ produce periodic density patterns; the density nodes of these patterns seed the eventual solitonification process~\cite{Banerjee2024, Bigelow2004}.
In our effective one-dimensional system the analytic Bogoliubov dispersion is $\omega_\kappa(q) = \sqrt{ \bar q^4 + 2 g \bar{\rho} \bar q^2}$, where $\bar q = q/ \sqrt{2n_0k_0}$, $n_0$ is the linear refractive index and $k_0$ is the laser wavevector.
Since $g$ is negative, the modes with $1/q>\xi = \sqrt{1/(g\bar \rho k_0)}$ have an imaginary frequency component whose value corresponds to the growth rate $\gamma$ of the mode.
The results shown in \cref{fig:collapse}c,d are from numerical calculations using the GPE.
The fragmentation of the ring is driven by the mode with the value $\eta$ that maximizes $\Im (\omega(q_\eta))$.
This mechanism is similar to the formation of bright soliton trains in elongated three-dimensional atomic condensates~\cite{Carr2004}.
The growth rate $\gamma$ decreases rapidly between $\kappa=1$ and $\kappa=9$, as shown in \cref{fig:collapse}d, indicating a delay of the collapse time $\hat\tau_c$ by more than an order of magnitude~\footnote{The GPE simulations for the case of $\kappa=0$, show a super-exponential growth of the density along $z$. Nevertheless, an exponential fit gives a lower bound of $\gamma=\qty{3}{mm^{-1}}$ leading to an overall effective delay of the collapse time by more than two orders of magnitude.} given that $\hat\tau_c \propto 1/\gamma_c$.

\begin{figure}
    \centering
    \includegraphics{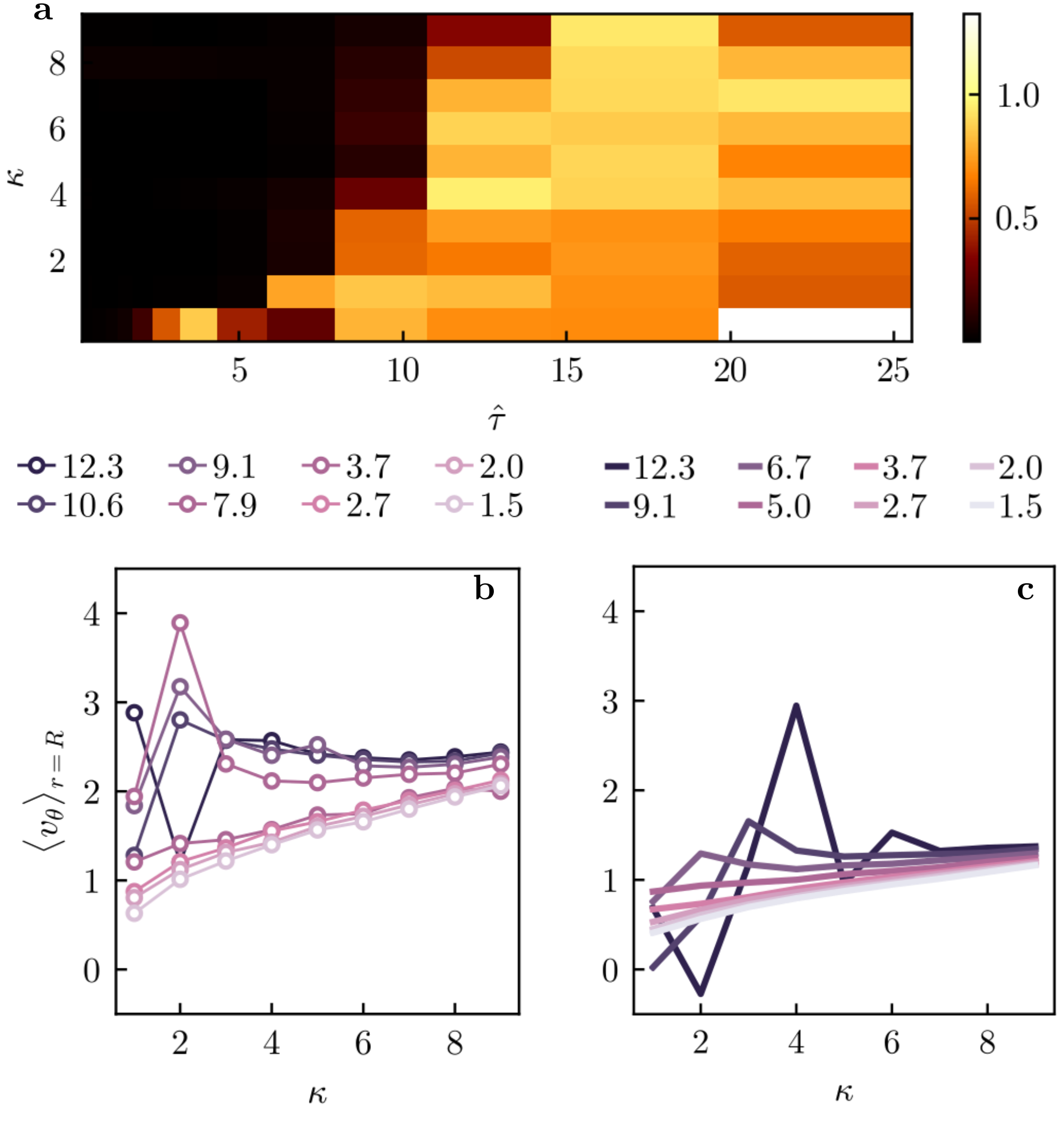}
    \caption{\textbf{a.} The amplitude of the dominant Bogoliubov mode normalised to $\bar\rho$ from GPE simulations. The time at which the ring collapses forming a solitonic pattern, can be roughly identified as the time when this quantity is $0.5$. The collapse time is delayed by an order of magnitude as $\kappa$ increases. The collapse dynamics affect the distribution of the azimuthal velocity shown in \textbf{b} (experiment) and \textbf{c} (GPE simulations). The different colours correspond to different times as in legend. For $\hat\tau < 9$ the azimuthal velocity scales linearly with $\kappa$, as expected for a quantized circulation; however, at higher values of $\hat\tau$ large variations appear when $\kappa$ is small, due to the complex dynamics leading to solitonification.}
    \label{fig:Vr_regions}
\end{figure}

\Cref{fig:Vr_regions}a shows the amplitude of the dominant Bogoliubov mode normalised to the mean density.
The collapse time $\hat\tau_c$ can be roughly identified as the line where the above ratio becomes $\approx 0.5$; this defines a collapse boundary that goes from $\hat\tau_c=1$--10 for $\kappa \leq 9$.
The increase of $\hat\tau_c$ is more prominent for small $\kappa$ and asymptotically saturates for $\kappa > 7$. 

Signatures of the collapse process can also be seen in the superfluid velocity which is proportional to the gradient of the phase of the BEC wavefunction $\psi({\bf r},\hat\tau)$.
The azimuthal velocity, $v_\mathrm{\theta} = \mathbf{v}\cdot \mathbf{\hat\theta}$ is expected to increase linearly with $\kappa$, which we observe only when $\hat\tau < 9$, see \cref{fig:Vr_regions}b,c.
Increasing $\hat\tau$ leads to large fluctuations of $v_\mathrm{\theta}$ that are more pronounced for small $\kappa$. The amplitude of these fluctuations persists for increasing $\kappa$ as $\hat\tau$ increases, and is in agreement with the collapse boundary in \cref{fig:Vr_regions}a. These fluctuations can be understood as necessarily opposing azimuthal flows that emerge during the collapse process and drive the fragmentation of the density.
At larger $\kappa$ the system recovers the linear scaling of $v_\mathrm{\theta}$ with $\kappa$.

In conclusion, using a quantum fluid of light as an analogue of a two-dimensional Bose-Einstein condensate, we demonstrate that topology can profoundly reshape the dynamics of an attractive BEC. While attractive condensates are intrinsically unstable and ultimately collapse through the growth of collective excitations, a quantized vortex introduces a competing flow that dramatically delays this process. Remarkably, increasing the vortex charge extends the collapse boundary by more than an order of magnitude in effective evolution time.

By directly measuring both density and phase, we reveal how the conservation of topological charge modifies the geometry and the velocity field of the condensate. The vortex-induced circulation counteracts the interaction-driven contraction, transforming the collapsing condensate into a ring whose radius, width, and stability are controlled by the vortex charge. The eventual breakdown of this state occurs through an azimuthal dynamical instability that fragments the ring into solitonic structures. The observed dynamics are in quantitative agreement with Gross-Pitaevskii simulations and a Bogoliubov analysis revealing the full excitation spectrum, which together identify the suppression of unstable azimuthal modes as the microscopic origin of the stabilization. 

More broadly, our results establish topology as a powerful control parameter for non-equilibrium quantum hydrodynamics, showing that topological defects can substantially alter the fate of unstable many-body systems. The mechanism unveiled here is generic and can be extended to multiple-defect configurations, vortex lattices, and higher-dimensional systems~\cite{Morris2024}, opening new directions for exploring the interplay between topology, collective excitations, and dynamical universal phenomena in quantum fluids.

\smallskip
\textbf{Acknowledgements.} 
We acknowledge fruitful discussions with I. Gnusov. We thank Paolo Cazzato for technical support.
We thank ECT$^\ast$ for support at the workshop ``Universal themes in Bose-Einstein Condensation'' during which part of this work was developed. F.D. thanks the Provincia autonoma di Trento for support.
This project was funded by PNRR MUR project: `National Quantum Science and Technology Institute' - NQSTI (PE0000023);
PNRR MUR project: ‘Integrated Infrastructure Initiative in Photonic and Quantum Sciences’ - I-PHOQS (IR0000016);
Quantum Optical Networks based on Exciton-polaritons - (Q-ONE) funding from the HORIZON-EIC-2022-PATHFINDER CHALLENGES EU programme under grant agreement No. 101115575;
Neuromorphic Polariton Accelerator - (PolArt) funding from the Horizon-EIC-2023-Pathfinder Open EU programme under grant agreement No.  101130304; 
the Italian Ministry of University and Research (MUR) under the granting scheme FIS 3 (grant number FIS-2024-04047).
Views and opinions expressed are however those of the author(s) only and do not necessarily reflect those of the European Union or European Innovation Council and SMEs Executive Agency (EISMEA). Neither the European Union nor the granting authority can be held responsible for them. 

\smallskip

\textbf{Competing interests.} 
The authors declare no competing interests.
\smallskip

\textbf{Data Availability.} 
The data of this study is available from the corresponding author upon reasonable request.
\smallskip

\textbf{Author contributions.} 
All authors contributed to discussions and editing of the manuscript.

\putbib
\end{bibunit}


\clearpage
\newpage

\begin{bibunit}

\onecolumngrid

\setcounter{equation}{0}
\setcounter{figure}{0}
\setcounter{table}{0}
\setcounter{section}{0}
\setcounter{page}{1}
\renewcommand{\theequation}{S\arabic{equation}}
\renewcommand{\thefigure}{S\arabic{figure}}

\section{Theoretical Analysis}

\lsection{Light field modelling and mapping to GPE}
A quantum fluid of light can be produced when a laser beam crosses a hot vapour~\cite{Carusotto2013,Glorieux2023}. 
Within the paraxial and slowly-varying envelope approximation~\cite{carusotto2014superfluid}, the electric field can be separated in a fast oscillating component and a slowly-varying envelope $\mathcal{E}({\bf r},z)$, where ${\bf r}$ is the position in the $xy$-plane and $z$ is the direction of propagation. 
In the regime where the light-vapour interactions become nonlinear, the envelope obeys the equation
\begin{equation}
 i \frac{\partial \mathcal{E}}{\partial z} = -\frac{\nabla^2_\perp \mathcal{E} }{2 n_0 k_0}-n_{2} \frac{c\epsilon_0k_0}{2n_0} \frac{\lvert \mathcal{E}\rvert ^2 \mathcal{E}}{1+\frac{\lvert \mathcal{E}\rvert ^2 }{\rho_{\mathrm{sat}}}}  -i \frac{\alpha}{2} \mathcal{E}
 \label{eq:optic}.
\end{equation}
Here, $k_0$ is the wavevector of the monochromatic beam, $\epsilon_0$ is the vacuum permittivity, $c$ is the speed of light, $n_0$ and $n_2$ are the linear and nonlinear parts of the refractive index, respectively, where $n_2$ is fixed by the third-order Kerr susceptibility of the atoms $\chi^{(3)}$ through the relation $\chi^{(3)} =  c \epsilon_0 n_2$. 
The term proportional to $n_2$ accounts for effective photon-photon interactions. 
The parameters $\rho_\mathrm{sat}$ and $\alpha$ quantify the nonlinear saturation and the absorption of the medium, respectively. 
In the experimental conditions, we can ignore the spatial variation of $n_0$, which would act as an effective external potential, and set $n_0=1$. 

\Cref{eq:optic} can be mapped onto a Gross-Pitaevskii-like equation (GPE)
\begin{equation}
 i \frac{\partial \psi} {\partial \tau}  = -\frac{\nabla^2_\perp \psi}{2n_0k_0} +  \frac{g \lvert \psi\rvert ^2 \psi}{1+\frac{\lvert \psi\rvert ^2}{\rho_\mathrm{sat}}} -i \frac{\alpha}{2} \psi,
 \label{eq:GPE}
\end{equation}
for the macroscopic wave function $\psi({\bf r},\tau)=\mathcal{E}({\bf r},z)$ of a two-dimensional condensate in a weakly interacting gas of bosons with density $\rho=|\psi|^2$.
The direction of propagation $z$ in \cref{eq:optic} plays the role of the effective time $\tau$ in the GPE. 
The quantity $g= - n_{2}c\epsilon_0k_0/(2n_0)$ is the coupling constant of the mean-field interaction. 
As such, each transverse plane of the cell at a fixed $z$ serves as a temporal snapshot representing the temporal evolution of a purely 2D system and, since $g$ is negative, the GPE describes an attractive BEC. 
We model the beam shape at the entrance of the cell with Gaussian wave function $\psi(r,0) = \sqrt{\rho_0} \exp(-r/\sigma_0^2)$, where the waist is $\sigma_0=210~\mu$m and $\rho_0$ is a parameter determining the laser intensity, $I_0=(1/2)c \epsilon_0 \rho_0$; we choose the value 
$I_\mathrm{0}: 1.23 \times 10^6\, \mathrm{W/m^2}$ in order to model a beam with same intensity as the one in the experiments. 
The topological charge $\kappa$ is  imprinted by multiplying the Gaussian by the phase factor $\exp(i \kappa \theta)$, with $\theta= \arctan(x,y)$, and we use different values of $\kappa$ up to $\kappa=9$.

In practice, we solve the GPE in dimensionless form, by introducing the characteristic nonlinear axial length $z_\mathrm{NL} = 1/(g \rho_0)$ and the effective healing length $\xi=\sqrt{z_\mathrm{NL}/k_0} = \sqrt{1/(g\rho_0 k_0)}$. 
We then use $\hat{\tau} = z/z_\mathrm{NL}$, and express the lengths in the $xy$ plane in units of $\xi$, which normalizes the wave function as $\hat{\psi} = \psi/\sqrt{\rho_0}$. 
The GPE becomes $
 i \partial \hat{\psi} / \partial \hat{\tau}  = [ - (1/2) \nabla^2_\perp + {\lvert \hat{\psi} \rvert ^2}/{(1+{\lvert \hat{\psi} \rvert ^2}/{\hat{\rho}_\mathrm{sat}}}) -i {\hat{\alpha}}/{2} ] \hat{\psi}
$, where $\hat{\rho}_\mathrm{sat} = \rho_\mathrm{sat}/\rho_0$ and $\hat{\alpha} = \alpha z_\mathrm{NL}$. 
We numerically solve this equation using the $4$th order Runge-Kutta algorithm of the XMDS2 package~\cite{DENNIS2013201} in a $512\times 512$ square box of size $L_\perp=\qty{0.4}{cm}$. 
It is worth stressing that, in the experiments, the light field is imaged at the end of the cell, which corresponds to the dimensionless time ${\hat\tau}_L = L/z_\mathrm{NL}$, with $L=\qty{7.5}{cm}$. 

\begin{figure}[t]
    \centering 
    \includegraphics[width=.6\textwidth]{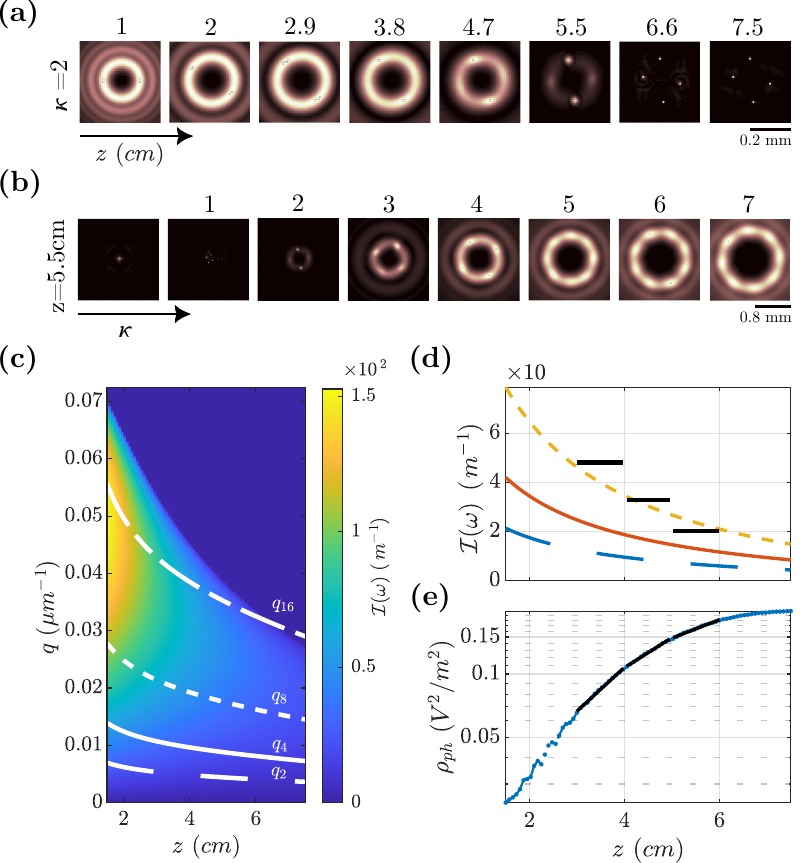}
    \caption{
    \textbf{Results of GPE numerical simulations for the ring formation and fragmentation.} 
    \textbf{a} Real-space density profiles of the light fluid along the propagation axis for $\kappa=2$. In 
    \textbf{b} the plots shows the density for different initially-imprinted circulation quanta $\kappa$, at $z = 5.5\,$cm. 
    In panel \textbf{c} we show the two-dimensional carpet plot of the imaginary part of the Bogoliubov spectra for the $\kappa=5$ case, calculated as in Eq.~\eqref{eq:Bogoliubov}. The dotted, solid and dashed lines follow the momentas 
    $q_\eta= \eta /2 R$ with $\eta=2,4,8,16$, respectively.  
    In \textbf{h}, the extracted imaginary parts $\mathcal{I}(\omega(q))$ along the momenta $q_\eta$ are plotted along the evolution, for the values $\eta=2,4,8$, as dashed blue, solid red and dotted yellow lines, respectively.
    The amplitude of the azimuthal density modulation is reported in panel \textbf{e}, along the longitudinal direction. The lines resulting from fitting with an exponential function in log-linear scale, over different regions($3-4$, $4-5$ and $5-6$cm), are reported in black colour. The numerical growth exponents, reported in panel \textbf{h}, extended over the corresponding fitting region, exhibit good agreement with the Bogoliubov prediction.
    }
    \label{fig:SM1}
\end{figure}

A typical example of simulation for a beam with $\kappa=2$ is reported in \cref{fig:SM1}a, for $g=-\qty{2.4} \times 10^{-7}{m/V^2}$, corresponding to a detuning of the laser $\Delta =-1.73 \ \mathrm{GHz}$ and $z_\mathrm{NL}=\qty{0.45}{cm}$. 
As saturation and absorption are expected to have a negligible influence on the mechanism driving the fragmentation of the ring into isolated spots discussed in the next section, we simplify the model by neglecting both effects, i.e., by taking $\hat{\rho}_\mathrm{sat}\to\infty$ and $\hat{\alpha}=0$.
Figure \ref{fig:SM1}a shows the formation of a ring with slowly varying radius and width, and its subsequent fragmentation in isolated peaks. 
Both the radius and the width can be extracted at each instant of the evolution by fitting the density at a given angle $\theta$ with a Gaussian function $\propto \exp(-(r-R)^2/2w^2)$, using $R$ and $w$ as fitting parameters, and then averaging over $\theta$. 
The predictions for the radius and the width at the exit of the cell (i.e. at $\tau=z=L$) as a function of $\hat\tau$ and for different values of $\kappa$, for the case without absorption and saturation, are shown as solid lines in Fig.~2. 

In \cref{fig:SM1}b we show snapshots of the intensity of the light at the same $z=5.5 \ \mathrm{cm}$ and for different values of $\kappa$. 

\lsection{Determination of the relation between $\Delta$ and $n_2$}
In the experiment, the nonlinear refractive index is varied using the detuning $\Delta$ which is directly controlled and measured. Although the $\chi_3$ value for Rb is known, its exact dependence on the atomic vapour parameters is not known. To establish a quantitative relationship between $\Delta$ and $n_2$, we combine experimental measurements with numerical simulations.
Experimentally, we vary the detuning $\Delta$ and the vorticity quanta parameter $\kappa$, and for each set of parameters we extract the ring diameter and width $w$ at the end of the evolution. These quantities provide the observables used for comparison with the numerical model.
On the numerical side, we compute extensive phase diagrams of the ring radius and width $w$ as functions of both $\Delta$ and $n_2$. We then assume a functional dependence of the form
\begin{equation}
n_2 = A e^{B\Delta},
\end{equation}
where $A$ and $B$ are fitting parameters. This relation defines a curve in the $(\Delta,n_2)$ parameter space, from which the simulated values of the ring radius and width can be extracted from the corresponding phase diagrams.
The coefficients $A$ and $B$ are determined by simultaneously fitting the experimentally measured radius and width $\sigma$ to the numerical predictions. In practice, the parameters are adjusted until the simulated observables reproduce the experimental values over the full range of detunings investigated.
The resulting comparison between experiment and simulations is shown in Fig.~2a and Fig.~2b of the main text. The best agreement is obtained for
\begin{equation}
A = 1.7 \times 10^{-6},
\qquad
B = 1.51,
\end{equation}
which provides the calibration curve relating the experimentally controlled detuning $\Delta$ to the effective nonlinear refractive index $n_2$.

\lsection{Ring fragmentation}
The fragmentation of the ring into isolated spots can be explained in terms of a modulational instability associated with the spectrum of Bogoliubov excitations on top of the slowly evolving ring-shaped condensate. 
Bogoliubov excitations corresponds to small oscillations around a stationary solution of the GPE. 
Since the dynamics of the ring is slow, we can assume that the Bogoliubov spectrum follows it adiabatically. 
The lowest energy excitations are the density and phase fluctuations along the ring, which are the analogue of long-wavelength phonons in a one-dimensional condensate; transverse excitations cost more energy because the width $w$ is much smaller than the length $2\pi R$. 
The periodic boundary conditions along the ring selects phonons with wavevector $q_\eta=\pm \eta/R$, with $\eta=1,2,3,\dots$, where $2\eta$ is the number of nodes. 
Counter propagating phonons with $q_\eta$ and $-q_\eta$ produce periodic density patterns with $2\eta$ maxima, which are indeed visible in the simulations, see \cref{fig:SM1}b. 
In good approximation, the growth of the most unstable Bogoliubov mode is exponential, as it can be seen in \cref{fig:SM1}e. 
A fit to its amplitude provides the growth rate, which varies with z, as a consequence of the slowly varying background. The growth rates calculated in this way are also plotted in Fig.~3d in the main text as diamonds with error bars.

To gain further insight, we can also estimate the growth rate as the imaginary frequency of the Bogoliubov modes extracted from their analytic dispersion relation in a uniform one-dimensional condensate \cite{Pitaevskii2016}, upon conversion in the notation of equation \eqref{eq:GPE}: 
\begin{equation}
    \omega_\kappa(q) = \sqrt{\left( \frac{q^2}{2n_0 k_0} \right)^2 + \frac{g \bar{\rho}}{n_0 k_0} q^2}\ .
    \label{eq:Bogoliubov}
\end{equation}
Here the condensate density $\bar{\rho}$ is an average density in the ring, that is, the density at $r=R$ averaged over the angle $\theta$ and divided by a numerical factor of order $1$ accounting for the transverse inhomogeneity. 
Since $R$ depends on $\kappa$, $\rho$ contains implicitly the dependence on the vortex charge.   
Following Refs.~\cite{PhysRevA.68.053611,Kramer2005}, we assume $\bar{\rho}$ to be half of the peak density, but this specific choice does not affect our main conclusions. 
What matters more is that, since in our case $g$ is negative, the modes with $q < \xi^{-1}$ have imaginary frequency, and the value of $\Im (\omega(q))$ corresponds to the growth rate of the Bogoliubov mode with wavevector $q$. 
The fragmentation of the ring occurs for the value $\eta$ such that $\Im (\omega(q_\eta))$ is the largest. 
To estimate the growth rates, we calculate $\Im (\omega(q,z))$ --- utilizing \cref{eq:Bogoliubov} and the numerically extracted $\bar{\rho}(z)$ ---, as shown in \cref{fig:SM1}c for the case $\kappa = 5$. 
To facilitate visualization, in panel c we show $\Im (\omega(q_\eta))$ along the propagation $z$, for the momenta $q_\eta = \eta / R(z)$ for $\eta = 1, 2, 4, 8$ --- with $R(z)$ the radius extracted numerically --- which are also plotted as white dotted, solid, and dashed lines in \cref{fig:SM1}(d).
Comparison with the numerically extracted growth rates (reported in panel e and discussed beforehand) is provided by reporting the latter as black horizontal lines in ~\cref{fig:SM1}d.

In Fig.~3c and d, we show the analytical predictions for the frequency $\omega(q)$ (dashed lines for the imaginary part and solid lines for the real part) for two values of $\kappa$, together with the corresponding growth rates for the modes with different $\eta$ (colored circles). 
The growth rates are also plotted in panel d of the same figure and compared with the largest growth rate extracted directly from the numerical simulations (diamonds). 
The agreement is very good and clearly supports the interpretation of the fragmentation as due to a modulation instability of Bogoliubov phonons in the attractive condensate. 
Evidence of this mechanism in the experimental observations can be seen for instance in panels a and b of Fig.~3. This type of dynamical instability is similar to the one observed in elongated atomic condensates~\cite{2017Nguyen} and theoretically studied in~\cite{2003Salasnich}.

\section{Local and global radial flow}
\begin{figure}
    \centering 
    \includegraphics{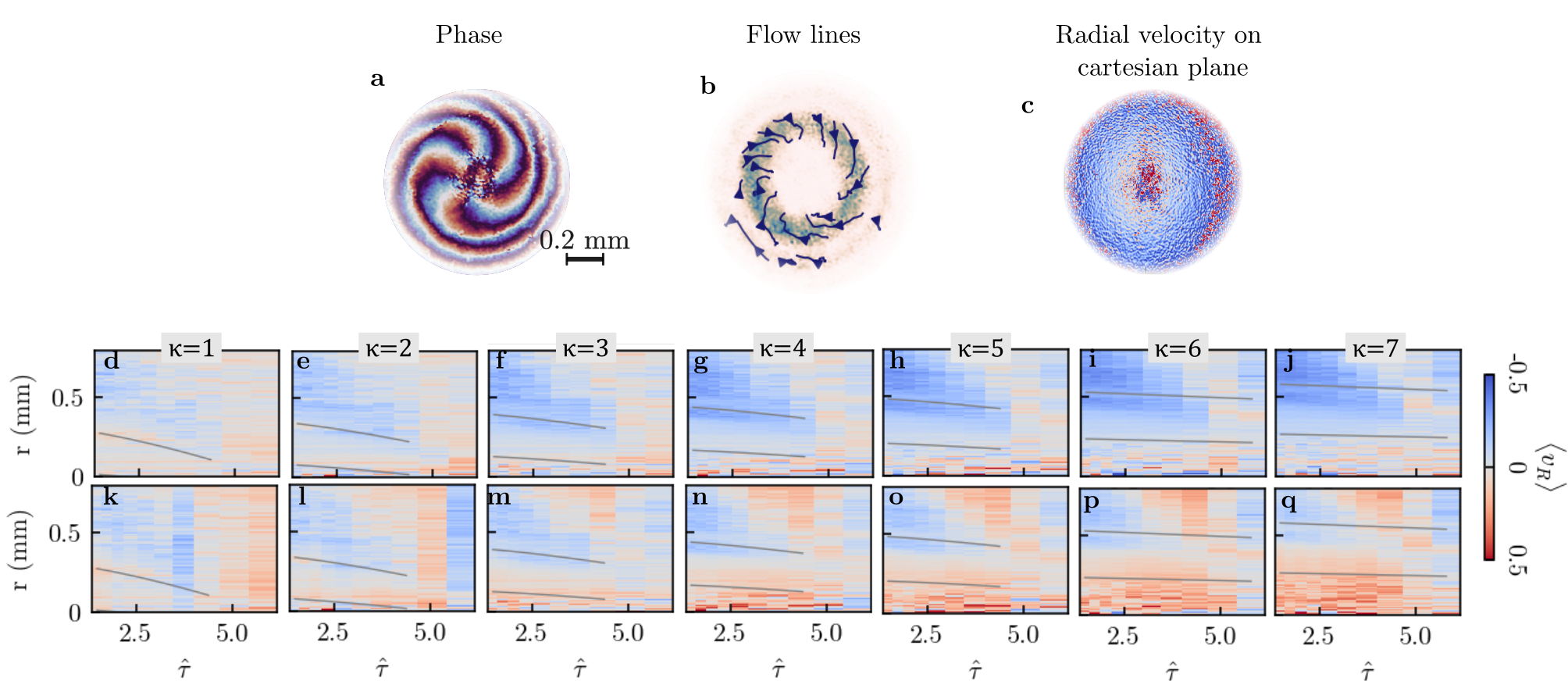}
    \caption{Velocity retrieval and local radial velocity. \textbf{a} From the hologram we retrieve the phase of the macroscopic wavefunction. \textbf{b} The gradient of the phase is proportional to the local velocity of the superfluid flow. \textbf{c} The flow is projected locally along the radial direction to retrieve the velocities. \textbf{d-q} For the radial flow, we report the radial velocity referred to the center of the ring in panels (d-j) and referred to the mean radius in panels (k-q). In both series the grey lines indicate the width of the ring. Setting the origin of the radial coordinate at the center of the ring (d-j) highlights an overall inward flow, translating the zero of the radial coordinate at the mean radius of the ring (k=q), highlights that the flow tends toward the mean radius (outward at small r and inward at large r).}
    \label{fig:flow_analysis}
\end{figure}

The addition of a vortex in the phase imparts both a radial and an azimuthal velocity on the BEC, which we can directly obtain from the measured spatial phase as $\nabla\phi (x, y)$. Experimentally, the phase $\phi$ of the macroscopic wavefunction is retrieved by interfering the output of the cell with a second beam with a flat phase at the detector, such that the phase difference of the interferogram corresponds to the phase of the beam passing through the cell, hence of the analogue condensate wavefunction $\psi$, see \cref{fig:flow_analysis}a.
The gradient along the cartesian axis of the image corresponds to the velocity along that axis $v_x=d\phi/dx$, $v_y=d\phi/dy$, therefore each point in the figure is mapped by a four-dimensional vector $(x,y,v_x,v_y)$. 
From here we can reconstruct both the radial and azimuthal velocities; here we discuss only the radial velocity.
It is natural to use polar coordinates to analyse the ring; first we project along the radial $\hat{r}$ direction setting the origin of the axis at the center of the ring, $v_r=v_x\cos(\theta)+v_y\sin(\theta)$. 
\Cref{fig:flow_analysis}b show the local radial velocity, respectively as $(x,y,v_r)$. 
Then the $(x,y)$-space is mapped to polar coordinates with the using the transformation $r=\sqrt{x^2+y^2}$ and $\theta=\arctan{y/x}$.

\Cref{fig:flow_analysis}(d-j) shows $\langle v_r \rangle_\theta$ as a function of $\hat\tau$. 
The sign of the flow is generally toward the center of the ring as expected from the attractive interaction, justifying the reduction in radius of the ring. 
When the ring is shrinking significantly a small outward flow appears in the inner part due to the conservation of the topological charge. 
By translating the origin of the radial direction $r_0$ at the mean radius of the ring $r_0=R$ we can obtain further information of the local radial flow within the thickness of the ring itself. 
\cref{fig:flow_analysis}(k-q) shows $\langle v_r \rangle_\theta$ referenced to the middle of the ring.
The grey lines indicate the width of the ring as obtained from the density profiles. 
In the inner part of the ring, the flow points outwards therefore toward the middle section of the ring and vice-versa for the outer part, where the flow is inward. This symmetry explains the reduction of the width, as discussed in the main text.

\section{Critical focusing power.}
\begin{figure}
    \centering
    \includegraphics[width=\textwidth]{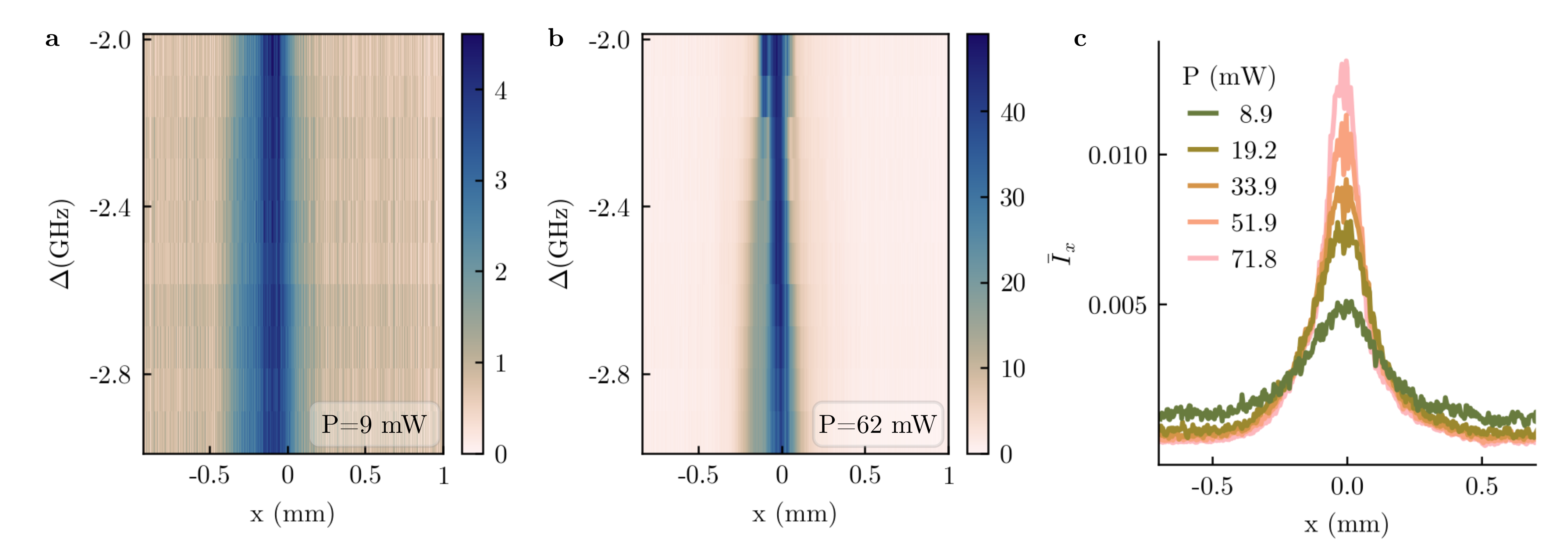}
    \caption{\textbf{a,b } Density profiles integrated along the y-direction and plotted as a function of $\Delta$. At $P\sim P_\mathrm{cr}$ (a), there is no focusing, while at $P\sim P_\mathrm{cr}$ (b), the system focuses until it collapses. Density profiles integrated along the x-direction, above $P_\mathrm{cr}$, showing that an increase in power leads to focusing.}
    \label{fig:gauss_power}
\end{figure}
For an attractive nonlinear medium, there is a minimum critical power for the onset of the focusing of a propagating beam defined as $P_\mathrm{cr}=\alpha \lambda^2/(4\pi n_0n_2)$,  where $\lambda$ is the laser wavelength, and $\alpha$ is a constant of the order of unity which depends on the beam shape. For a gaussian beam $\alpha\sim1.8$ \cite{Fibich2000}, which sets our critical power at \qty{8.7}{mW} for $n_2=1\times10^{-11}$. Here we use a gaussian beam and scan $\Delta$ for $P\sim P_\mathrm{cr}$ and $P>P_\mathrm{cr}$, to observe the onset of the focusing dynamic. \Cref{fig:gauss_power}a,b shows the density profiles integrated along the x-axis. For $P\sim P_\mathrm{cr}$ there is no focusing and the size of the beam is in good approximation constant, on the contrary, starting at $P\sim8P_\mathrm{cr}$ the beam is focusing aggressively over the same range of detuning up to fragmentation, which takes place around $\Delta=-2.4\,\mathrm{GHz}$ ($\hat\tau=3.4$). 
Equivalently, above $P_\mathrm{cr}$, increasing the power leads to an increase in the focusing for a fix $\Delta$, as shown by the normalized density profiles in  \cref{fig:gauss_power}c.
With the vortex of charge $\kappa$ the critical focusing power scales as $\kappa^{3/2}$ for large $\kappa$, as discussed in~\cite{vuong2006,kruglov1992}.

\section{Absorption and fragmentation}
\begin{figure}
\centering
\includegraphics{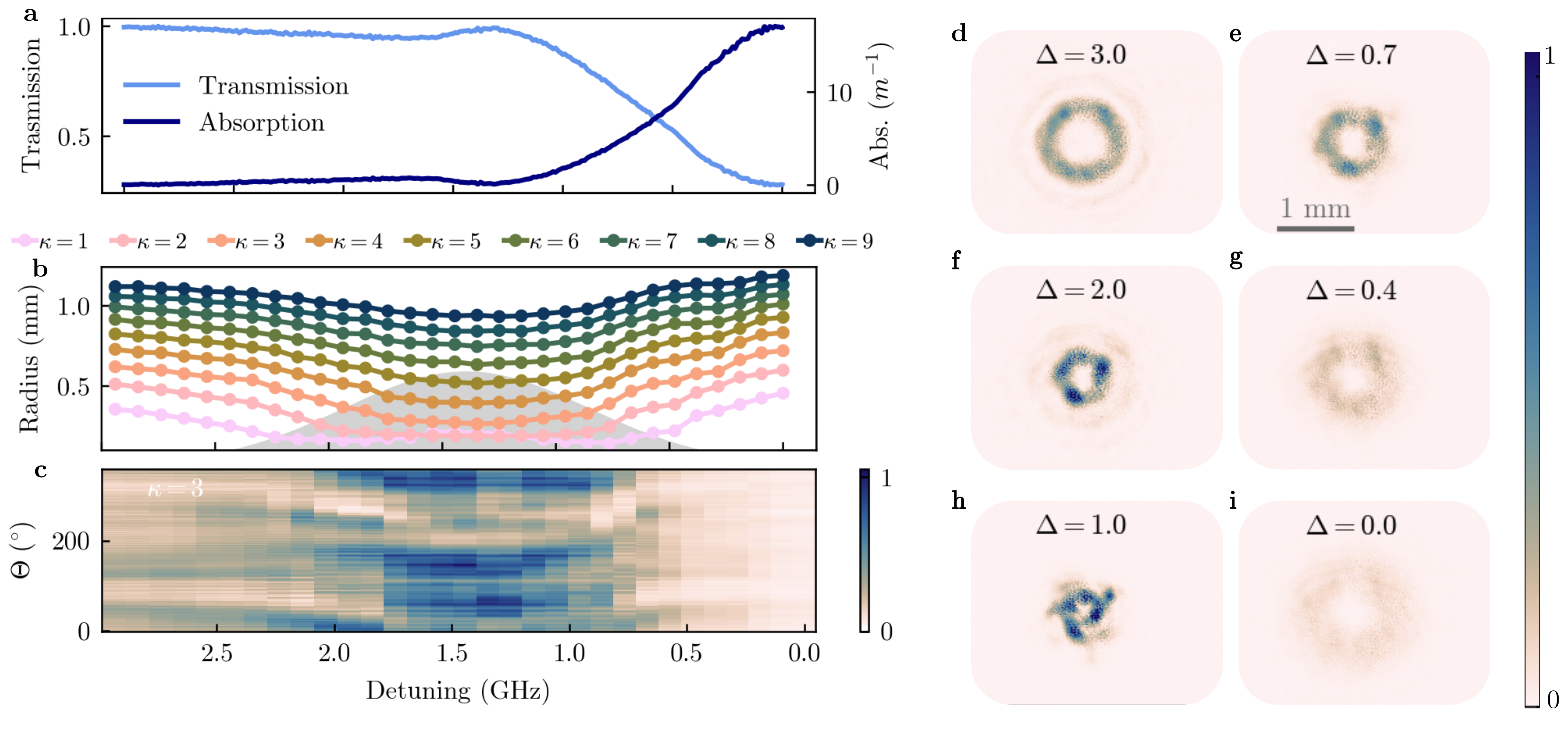}
\caption{Effects of absorption on the dynamics. \textbf{a} Normalized experimental transmission and derived absorption coefficient, for reference. \textbf{b} Radii decrease with increasing nonlinearity until absorption takes over and slows down the dynamics, the shaded gray area indicates the region where the beam is completely fragmented. \textbf{c}  Azimuthal density along the ring, showing the fragmentation and rotation of the ring and the inversion due absorption. 
\textbf{d--i} Density pictures for different detunings, showing the ring fragmentation and recomposition. The colorscale for the density has been renormalized to account for the reduced intensity at the camera. }
\label{Fig:Absorption}
\end{figure}
A strong advantage of quantum fluids of light in hot atomic vapour is the possibility to work in a regime where absorption is negligible and the dynamics can be considered conservative. This regime is achieved at $\Delta$ greater than the effective linewidth of the system, that for a thermal gas is dominated by the Doppler broadening. In our work, we enforce this condition by choosing the range on the blue side of the transition $\Delta=1.5\text{--}3\,\mathrm{GHz}$ where the transmission $T>99\%$ and the corresponding absorption value $\alpha<\,$\qty{0.1}{m^{-1}}. In this regime, reducing the detuning, while keeping the other quantities constant, increases the nonlinearity and consequently the effective time. The reduction in the radii of the rings and the emergence of fragmentation as the system enters the solitonic regime discussed in the main text are a direct consequence of this. Approaching the atomic resonance with the gas at 120$^\circ\mathrm{C}$, and $P=85\,\mathrm{mW}$ for the laser beam, absorption increases up to 75\%, see \cref{Fig:Absorption}a. Starting from $\Delta\sim1.3\,\mathrm{GHz}$, where absorption starts to become significant, the dynamics begin to reverse. \Cref{Fig:Absorption}b is an extension of Fig.\,2a in the main text. Here, the radii are defined as the center of mass of the system along the radial coordinate, to account for the large inhomogeneities in the regime of the modulation instability and solitonic formation.  While below the turning point defined by absorption the ring size decreases, above the turning point the radii increase again.

Even more striking is the inversion observed in the azimuthal dynamics. By integrating the density radially in a \qty{100}{\mu m}-region around the radius, it is possible to extract a one-dimensional approximation of the azimuthal density and see how it changes with the detuning \cref{Fig:Absorption}b. The initial azimuthal density has only small perturbations due to small inhomogeneities in the beam, however as the detuning increases a density modulation appears with high-density peaks forming, as shown in \cref{Fig:Absorption}\textbf{c}. While in the numerical simulation such peaks are distributed periodically, in the experiment the small inhomogeneity can initially seed the peak formation; this difference does not alter the dynamics. The peak positions are shifted by \qty{60}{\degree} as the detuning decreases, which signifies the rotation of the ring initiated by the vortex. At $\lvert \Delta\rvert<$\qty{1.3}{GHz}, the shift decreases again and the peaks due to the density modulation and soliton formation rotate back to their original position around $\Delta=-0.5$. At the same time, the density gets significantly depleted due to the increased absorption.
\Cref{Fig:Absorption}e--j show the different phases the beam undergoes as the initial ring collapses, reduces its size and fragments. As absorption increases, the fragmentation disappears, the ring becomes homogeneous again and comparable in size to the case at large detuning.

\section{Off-centred vortex}
\begin{figure}
     \centering
     \includegraphics[width=0.5\linewidth]{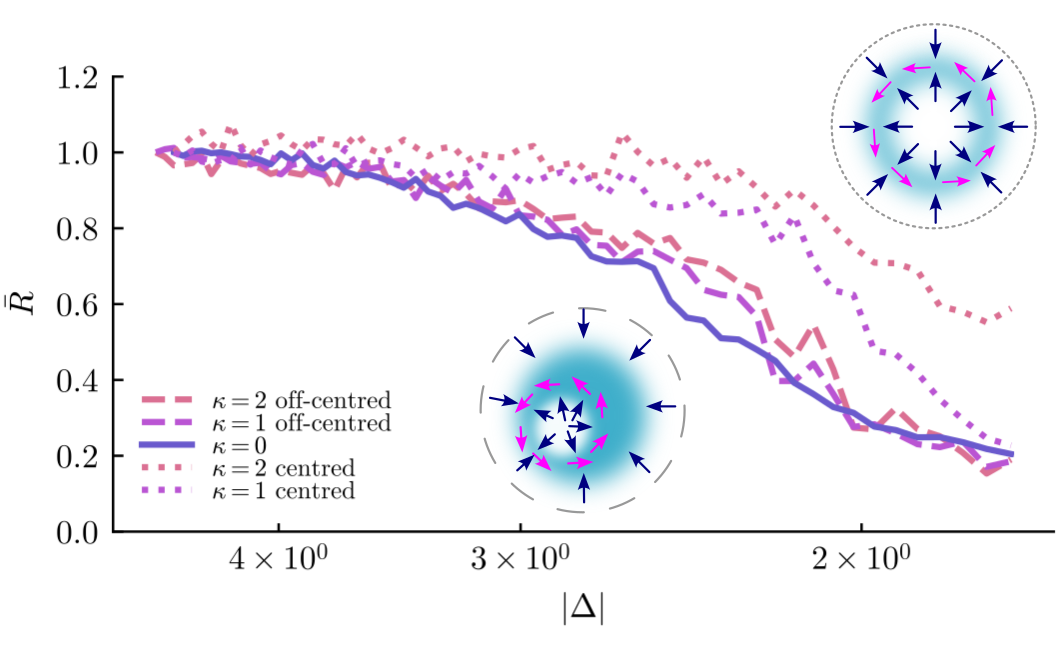}
     \caption{Normalized radius $\bar R$ as a function of $\Delta$, with centred(dotted) and off-centred (dashed) vortex, for $\kappa=0/2$indicated in the legend. The inset show a cartoon picture of vorticity and interactions for the case with centred and off-centred vortex, respectively top-right and bottom-left.}
     \label{fig:off-cent}
\end{figure}

As the interactions have cylindrical symmetry around the peak density of the fluid, a vortex at its center is most effective at dynamically stabilising it. 
Indeed, when imprinting the vortex closer to the edge of the fluid, the collapse remains practically unchanged with respect to the $\kappa=0$ case.
\Cref{fig:off-cent} shows the average of gaussian fits to the integral of the density profiles over x and y as a proxy of the size of the BEC.

For a centred vortex the final size of the fluid is larger than the $\kappa=0$ case, similarly to before. 
On the contrary, when the defect is off-centre, the decrease in size starts for the same $n_2$ independently for $\kappa=1,2$ and it always reaches the same value as $\kappa=0$, confirming that the stabilization is less efficient when the vortex is not centred.

\putbib
\end{bibunit}


\begin{thebibliography}{27}%
\makeatletter
\providecommand \@ifxundefined [1]{%
 \@ifx{#1\undefined}
}%
\providecommand \@ifnum [1]{%
 \ifnum #1\expandafter \@firstoftwo
 \else \expandafter \@secondoftwo
 \fi
}%
\providecommand \@ifx [1]{%
 \ifx #1\expandafter \@firstoftwo
 \else \expandafter \@secondoftwo
 \fi
}%
\providecommand \natexlab [1]{#1}%
\providecommand \enquote  [1]{``#1''}%
\providecommand \bibnamefont  [1]{#1}%
\providecommand \bibfnamefont [1]{#1}%
\providecommand \citenamefont [1]{#1}%
\providecommand \href@noop [0]{\@secondoftwo}%
\providecommand \href [0]{\begingroup \@sanitize@url \@href}%
\providecommand \@href[1]{\@@startlink{#1}\@@href}%
\providecommand \@@href[1]{\endgroup#1\@@endlink}%
\providecommand \@sanitize@url [0]{\catcode `\\12\catcode `\$12\catcode
  `\&12\catcode `\#12\catcode `\^12\catcode `\_12\catcode `\%12\relax}%
\providecommand \@@startlink[1]{}%
\providecommand \@@endlink[0]{}%
\providecommand \url  [0]{\begingroup\@sanitize@url \@url }%
\providecommand \@url [1]{\endgroup\@href {#1}{\urlprefix }}%
\providecommand \urlprefix  [0]{URL }%
\providecommand \Eprint [0]{\href }%
\providecommand \doibase [0]{https://doi.org/}%
\providecommand \selectlanguage [0]{\@gobble}%
\providecommand \bibinfo  [0]{\@secondoftwo}%
\providecommand \bibfield  [0]{\@secondoftwo}%
\providecommand \translation [1]{[#1]}%
\providecommand \BibitemOpen [0]{}%
\providecommand \bibitemStop [0]{}%
\providecommand \bibitemNoStop [0]{.\EOS\space}%
\providecommand \EOS [0]{\spacefactor3000\relax}%
\providecommand \BibitemShut  [1]{\csname bibitem#1\endcsname}%
\let\auto@bib@innerbib\@empty
\bibitem [{\citenamefont {Mermin}(1979)}]{Mermin1979}%
  \BibitemOpen
  \bibfield  {author} {\bibinfo {author} {\bibfnamefont {N.~D.}\ \bibnamefont
  {Mermin}},\ }\href {https://doi.org/10.1103/RevModPhys.51.591} {\bibfield
  {journal} {\bibinfo  {journal} {Rev. Mod. Phys.}\ }\textbf {\bibinfo {volume}
  {51}},\ \bibinfo {pages} {591} (\bibinfo {year} {1979})}\BibitemShut
  {NoStop}%
\bibitem [{\citenamefont {Nelson}(2002)}]{nelson2002}%
  \BibitemOpen
  \bibfield  {author} {\bibinfo {author} {\bibfnamefont {D.~R.}\ \bibnamefont
  {Nelson}},\ }\href@noop {} {\emph {\bibinfo {title} {Defects and geometry in
  condensed matter physics}}}\ (\bibinfo  {publisher} {Cambridge University
  Press},\ \bibinfo {year} {2002})\BibitemShut {NoStop}%
\bibitem [{\citenamefont {Kibble}(1976)}]{Kibble1976}%
  \BibitemOpen
  \bibfield  {author} {\bibinfo {author} {\bibfnamefont {T.~W.~B.}\
  \bibnamefont {Kibble}},\ }\href {https://doi.org/10.1088/0305-4470/9/8/029}
  {\bibfield  {journal} {\bibinfo  {journal} {Journal of Physics A:
  Mathematical and General}\ }\textbf {\bibinfo {volume} {9}},\ \bibinfo
  {pages} {1387} (\bibinfo {year} {1976})}\BibitemShut {NoStop}%
\bibitem [{\citenamefont {Arda\v{s}eva}\ and\ \citenamefont
  {Doostmohammadi}(2022)}]{Ardaseva2022}%
  \BibitemOpen
  \bibfield  {author} {\bibinfo {author} {\bibfnamefont {A.}~\bibnamefont
  {Arda\v{s}eva}}\ and\ \bibinfo {author} {\bibfnamefont {A.}~\bibnamefont
  {Doostmohammadi}},\ }\href {https://doi.org/10.1038/s42254-022-00469-9}
  {\bibfield  {journal} {\bibinfo  {journal} {Nat. Rev. Phys.}\ }\textbf
  {\bibinfo {volume} {4}} (\bibinfo {year} {2022})}\BibitemShut {NoStop}%
\bibitem [{\citenamefont {Shankar}\ \emph {et~al.}(2022)\citenamefont
  {Shankar}, \citenamefont {Souslov}, \citenamefont {Bowick}, \citenamefont
  {Marchetti},\ and\ \citenamefont {Vitelli}}]{Shankar2022}%
  \BibitemOpen
  \bibfield  {author} {\bibinfo {author} {\bibfnamefont {S.}~\bibnamefont
  {Shankar}}, \bibinfo {author} {\bibfnamefont {A.}~\bibnamefont {Souslov}},
  \bibinfo {author} {\bibfnamefont {M.~J.}\ \bibnamefont {Bowick}}, \bibinfo
  {author} {\bibfnamefont {M.~C.}\ \bibnamefont {Marchetti}},\ and\ \bibinfo
  {author} {\bibfnamefont {V.}~\bibnamefont {Vitelli}},\ }\href
  {https://doi.org/10.1038/s42254-022-00445-3} {\bibfield  {journal} {\bibinfo
  {journal} {Nat. Rev. Phys.}\ }\textbf {\bibinfo {volume} {4}} (\bibinfo
  {year} {2022})}\BibitemShut {NoStop}%
\bibitem [{\citenamefont {Teo}\ and\ \citenamefont
  {Hughes}(2017)}]{Jeffrey2017}%
  \BibitemOpen
  \bibfield  {author} {\bibinfo {author} {\bibfnamefont {J.~C.}\ \bibnamefont
  {Teo}}\ and\ \bibinfo {author} {\bibfnamefont {T.~L.}\ \bibnamefont
  {Hughes}},\ }\href
  {https://doi.org/https://doi.org/10.1146/annurev-conmatphys-031016-025154}
  {\bibfield  {journal} {\bibinfo  {journal} {Annual Review of Condensed Matter
  Physics}\ }\textbf {\bibinfo {volume} {8}},\ \bibinfo {pages} {211} (\bibinfo
  {year} {2017})}\BibitemShut {NoStop}%
\bibitem [{\citenamefont {Sackett}\ \emph {et~al.}(1998)\citenamefont
  {Sackett}, \citenamefont {Stoof},\ and\ \citenamefont {Hulet}}]{Sackett1998}%
  \BibitemOpen
  \bibfield  {author} {\bibinfo {author} {\bibfnamefont {C.~A.}\ \bibnamefont
  {Sackett}}, \bibinfo {author} {\bibfnamefont {H.~T.~C.}\ \bibnamefont
  {Stoof}},\ and\ \bibinfo {author} {\bibfnamefont {R.~G.}\ \bibnamefont
  {Hulet}},\ }\href {https://doi.org/10.1103/PhysRevLett.80.2031} {\bibfield
  {journal} {\bibinfo  {journal} {Phys. Rev. Lett.}\ }\textbf {\bibinfo
  {volume} {80}},\ \bibinfo {pages} {2031} (\bibinfo {year}
  {1998})}\BibitemShut {NoStop}%
\bibitem [{\citenamefont {Roberts}\ \emph {et~al.}(2001)\citenamefont
  {Roberts}, \citenamefont {Claussen}, \citenamefont {Cornish}, \citenamefont
  {Donley}, \citenamefont {Cornell},\ and\ \citenamefont
  {Wieman}}]{Roberts2001}%
  \BibitemOpen
  \bibfield  {author} {\bibinfo {author} {\bibfnamefont {J.~L.}\ \bibnamefont
  {Roberts}}, \bibinfo {author} {\bibfnamefont {N.~R.}\ \bibnamefont
  {Claussen}}, \bibinfo {author} {\bibfnamefont {S.~L.}\ \bibnamefont
  {Cornish}}, \bibinfo {author} {\bibfnamefont {E.~A.}\ \bibnamefont {Donley}},
  \bibinfo {author} {\bibfnamefont {E.~A.}\ \bibnamefont {Cornell}},\ and\
  \bibinfo {author} {\bibfnamefont {C.~E.}\ \bibnamefont {Wieman}},\ }\href
  {https://doi.org/10.1103/PhysRevLett.86.4211} {\bibfield  {journal} {\bibinfo
   {journal} {Phys. Rev. Lett.}\ }\textbf {\bibinfo {volume} {86}},\ \bibinfo
  {pages} {4211} (\bibinfo {year} {2001})}\BibitemShut {NoStop}%
\bibitem [{\citenamefont {Cornish}\ \emph {et~al.}(2006)\citenamefont
  {Cornish}, \citenamefont {Thompson},\ and\ \citenamefont
  {Wieman}}]{Cornish2006}%
  \BibitemOpen
  \bibfield  {author} {\bibinfo {author} {\bibfnamefont {S.~L.}\ \bibnamefont
  {Cornish}}, \bibinfo {author} {\bibfnamefont {S.~T.}\ \bibnamefont
  {Thompson}},\ and\ \bibinfo {author} {\bibfnamefont {C.~E.}\ \bibnamefont
  {Wieman}},\ }\href {https://doi.org/10.1103/PhysRevLett.96.170401} {\bibfield
   {journal} {\bibinfo  {journal} {Phys. Rev. Lett.}\ }\textbf {\bibinfo
  {volume} {96}},\ \bibinfo {pages} {170401} (\bibinfo {year}
  {2006})}\BibitemShut {NoStop}%
\bibitem [{\citenamefont {Banerjee}\ \emph {et~al.}(2025)\citenamefont
  {Banerjee}, \citenamefont {Zhou}, \citenamefont {Tiwari}, \citenamefont
  {Tamura}, \citenamefont {Li}, \citenamefont {Kevrekidis}, \citenamefont
  {Mistakidis}, \citenamefont {Walther},\ and\ \citenamefont
  {Hung}}]{Banerjee2024}%
  \BibitemOpen
  \bibfield  {author} {\bibinfo {author} {\bibfnamefont {S.}~\bibnamefont
  {Banerjee}}, \bibinfo {author} {\bibfnamefont {K.}~\bibnamefont {Zhou}},
  \bibinfo {author} {\bibfnamefont {S.~K.}\ \bibnamefont {Tiwari}}, \bibinfo
  {author} {\bibfnamefont {H.}~\bibnamefont {Tamura}}, \bibinfo {author}
  {\bibfnamefont {R.}~\bibnamefont {Li}}, \bibinfo {author} {\bibfnamefont
  {P.}~\bibnamefont {Kevrekidis}}, \bibinfo {author} {\bibfnamefont {S.~I.}\
  \bibnamefont {Mistakidis}}, \bibinfo {author} {\bibfnamefont
  {V.}~\bibnamefont {Walther}},\ and\ \bibinfo {author} {\bibfnamefont {C.-L.}\
  \bibnamefont {Hung}},\ }\href {https://doi.org/10.1103/c6wx-zc9x} {\bibfield
  {journal} {\bibinfo  {journal} {Phys. Rev. Lett.}\ }\textbf {\bibinfo
  {volume} {135}},\ \bibinfo {pages} {073401} (\bibinfo {year}
  {2025})}\BibitemShut {NoStop}%
\bibitem [{\citenamefont {Modugno}\ \emph {et~al.}(2002)\citenamefont
  {Modugno}, \citenamefont {Roati}, \citenamefont {Riboli}, \citenamefont
  {Ferlaino}, \citenamefont {Brecha},\ and\ \citenamefont
  {Inguscio}}]{Modugno2002}%
  \BibitemOpen
  \bibfield  {author} {\bibinfo {author} {\bibfnamefont {G.}~\bibnamefont
  {Modugno}}, \bibinfo {author} {\bibfnamefont {G.}~\bibnamefont {Roati}},
  \bibinfo {author} {\bibfnamefont {F.}~\bibnamefont {Riboli}}, \bibinfo
  {author} {\bibfnamefont {F.}~\bibnamefont {Ferlaino}}, \bibinfo {author}
  {\bibfnamefont {R.~J.}\ \bibnamefont {Brecha}},\ and\ \bibinfo {author}
  {\bibfnamefont {M.}~\bibnamefont {Inguscio}},\ }\href
  {https://doi.org/10.1126/science.1077386} {\bibfield  {journal} {\bibinfo
  {journal} {Science}\ }\textbf {\bibinfo {volume} {297}},\ \bibinfo {pages}
  {2240} (\bibinfo {year} {2002})}\BibitemShut {NoStop}%
\bibitem [{\citenamefont {Dalfovo}\ and\ \citenamefont
  {Stringari}(1996)}]{Dalfovo1996}%
  \BibitemOpen
  \bibfield  {author} {\bibinfo {author} {\bibfnamefont {F.}~\bibnamefont
  {Dalfovo}}\ and\ \bibinfo {author} {\bibfnamefont {S.}~\bibnamefont
  {Stringari}},\ }\href {https://doi.org/10.1103/PhysRevA.53.2477} {\bibfield
  {journal} {\bibinfo  {journal} {Phys. Rev. A}\ }\textbf {\bibinfo {volume}
  {53}},\ \bibinfo {pages} {2477} (\bibinfo {year} {1996})}\BibitemShut
  {NoStop}%
\bibitem [{\citenamefont {Nesti}\ and\ \citenamefont
  {Pezzè}(2026)}]{Nesti_2026}%
  \BibitemOpen
  \bibfield  {author} {\bibinfo {author} {\bibfnamefont {G.}~\bibnamefont
  {Nesti}}\ and\ \bibinfo {author} {\bibfnamefont {L.}~\bibnamefont {Pezzè}},\
  }\bibfield  {journal} {\bibinfo  {journal} {Physical Review A}\ }\textbf
  {\bibinfo {volume} {113}},\ \href {https://doi.org/10.1103/tb58-33h6}
  {10.1103/tb58-33h6} (\bibinfo {year} {2026})\BibitemShut {NoStop}%
\bibitem [{\citenamefont {Bigelow}\ \emph {et~al.}(2004)\citenamefont
  {Bigelow}, \citenamefont {Zerom},\ and\ \citenamefont {Boyd}}]{Bigelow2004}%
  \BibitemOpen
  \bibfield  {author} {\bibinfo {author} {\bibfnamefont {M.~S.}\ \bibnamefont
  {Bigelow}}, \bibinfo {author} {\bibfnamefont {P.}~\bibnamefont {Zerom}},\
  and\ \bibinfo {author} {\bibfnamefont {R.~W.}\ \bibnamefont {Boyd}},\ }\href
  {https://doi.org/10.1103/PhysRevLett.92.083902} {\bibfield  {journal}
  {\bibinfo  {journal} {Phys. Rev. Lett.}\ }\textbf {\bibinfo {volume} {92}},\
  \bibinfo {pages} {083902} (\bibinfo {year} {2004})}\BibitemShut {NoStop}%
\bibitem [{\citenamefont {Petrov}\ \emph {et~al.}(1998)\citenamefont {Petrov},
  \citenamefont {Torner}, \citenamefont {Martorell}, \citenamefont {Vilaseca},
  \citenamefont {Torres},\ and\ \citenamefont {Cojocaru}}]{Petrov1998}%
  \BibitemOpen
  \bibfield  {author} {\bibinfo {author} {\bibfnamefont {D.~V.}\ \bibnamefont
  {Petrov}}, \bibinfo {author} {\bibfnamefont {L.}~\bibnamefont {Torner}},
  \bibinfo {author} {\bibfnamefont {J.}~\bibnamefont {Martorell}}, \bibinfo
  {author} {\bibfnamefont {R.}~\bibnamefont {Vilaseca}}, \bibinfo {author}
  {\bibfnamefont {J.~P.}\ \bibnamefont {Torres}},\ and\ \bibinfo {author}
  {\bibfnamefont {C.}~\bibnamefont {Cojocaru}},\ }\href
  {https://doi.org/10.1364/OL.23.001444} {\bibfield  {journal} {\bibinfo
  {journal} {Opt. Lett.}\ }\textbf {\bibinfo {volume} {23}},\ \bibinfo {pages}
  {1444} (\bibinfo {year} {1998})}\BibitemShut {NoStop}%
\bibitem [{\citenamefont {Vuong}\ \emph {et~al.}(2006)\citenamefont {Vuong},
  \citenamefont {Grow}, \citenamefont {Ishaaya}, \citenamefont {Gaeta},
  \citenamefont {'t~Hooft}, \citenamefont {Eliel},\ and\ \citenamefont
  {Fibich}}]{vuong2006}%
  \BibitemOpen
  \bibfield  {author} {\bibinfo {author} {\bibfnamefont {L.~T.}\ \bibnamefont
  {Vuong}}, \bibinfo {author} {\bibfnamefont {T.~D.}\ \bibnamefont {Grow}},
  \bibinfo {author} {\bibfnamefont {A.}~\bibnamefont {Ishaaya}}, \bibinfo
  {author} {\bibfnamefont {A.~L.}\ \bibnamefont {Gaeta}}, \bibinfo {author}
  {\bibfnamefont {G.~W.}\ \bibnamefont {'t~Hooft}}, \bibinfo {author}
  {\bibfnamefont {E.~R.}\ \bibnamefont {Eliel}},\ and\ \bibinfo {author}
  {\bibfnamefont {G.}~\bibnamefont {Fibich}},\ }\href
  {https://doi.org/10.1103/PhysRevLett.96.133901} {\bibfield  {journal}
  {\bibinfo  {journal} {Phys. Rev. Lett.}\ }\textbf {\bibinfo {volume} {96}},\
  \bibinfo {pages} {133901} (\bibinfo {year} {2006})}\BibitemShut {NoStop}%
\bibitem [{\citenamefont {Donley}\ \emph {et~al.}(2001)\citenamefont {Donley},
  \citenamefont {Claussen}, \citenamefont {Cornish}, \citenamefont {Roberts},
  \citenamefont {Cornell},\ and\ \citenamefont {Wieman}}]{Donley2001}%
  \BibitemOpen
  \bibfield  {author} {\bibinfo {author} {\bibfnamefont {E.~A.}\ \bibnamefont
  {Donley}}, \bibinfo {author} {\bibfnamefont {N.~R.}\ \bibnamefont
  {Claussen}}, \bibinfo {author} {\bibfnamefont {S.~L.}\ \bibnamefont
  {Cornish}}, \bibinfo {author} {\bibfnamefont {J.~L.}\ \bibnamefont
  {Roberts}}, \bibinfo {author} {\bibfnamefont {E.~A.}\ \bibnamefont
  {Cornell}},\ and\ \bibinfo {author} {\bibfnamefont {C.~E.}\ \bibnamefont
  {Wieman}},\ }\href {https://doi.org/https://doi.org/10.1038/35085500}
  {\bibfield  {journal} {\bibinfo  {journal} {Nature}\ }\textbf {\bibinfo
  {volume} {412}},\ \bibinfo {pages} {295} (\bibinfo {year}
  {2001})}\BibitemShut {NoStop}%
\bibitem [{\citenamefont {Westerberg}\ \emph {et~al.}(2018)\citenamefont
  {Westerberg}, \citenamefont {Wilson}, \citenamefont {Duncan}, \citenamefont
  {Faccio}, \citenamefont {Wright}, \citenamefont {\"Ohberg},\ and\
  \citenamefont {Valiente}}]{westerberg.PhysRevA.98.053835}%
  \BibitemOpen
  \bibfield  {author} {\bibinfo {author} {\bibfnamefont {N.}~\bibnamefont
  {Westerberg}}, \bibinfo {author} {\bibfnamefont {K.~E.}\ \bibnamefont
  {Wilson}}, \bibinfo {author} {\bibfnamefont {C.~W.}\ \bibnamefont {Duncan}},
  \bibinfo {author} {\bibfnamefont {D.}~\bibnamefont {Faccio}}, \bibinfo
  {author} {\bibfnamefont {E.~M.}\ \bibnamefont {Wright}}, \bibinfo {author}
  {\bibfnamefont {P.}~\bibnamefont {\"Ohberg}},\ and\ \bibinfo {author}
  {\bibfnamefont {M.}~\bibnamefont {Valiente}},\ }\href
  {https://doi.org/10.1103/PhysRevA.98.053835} {\bibfield  {journal} {\bibinfo
  {journal} {Phys. Rev. A}\ }\textbf {\bibinfo {volume} {98}},\ \bibinfo
  {pages} {053835} (\bibinfo {year} {2018})}\BibitemShut {NoStop}%
\bibitem [{\citenamefont {Nguyen}\ \emph {et~al.}(2017)\citenamefont {Nguyen},
  \citenamefont {Luo},\ and\ \citenamefont {Hulet}}]{2017Nguyen}%
  \BibitemOpen
  \bibfield  {author} {\bibinfo {author} {\bibfnamefont {J.~H.~V.}\
  \bibnamefont {Nguyen}}, \bibinfo {author} {\bibfnamefont {D.}~\bibnamefont
  {Luo}},\ and\ \bibinfo {author} {\bibfnamefont {R.~G.}\ \bibnamefont
  {Hulet}},\ }\href {https://doi.org/10.1126/science.aal3220} {\bibfield
  {journal} {\bibinfo  {journal} {Science}\ }\textbf {\bibinfo {volume}
  {356}},\ \bibinfo {pages} {422} (\bibinfo {year} {2017})}\BibitemShut
  {NoStop}%
\bibitem [{\citenamefont {Salasnich}\ \emph {et~al.}(2003)\citenamefont
  {Salasnich}, \citenamefont {Parola},\ and\ \citenamefont
  {Reatto}}]{2003Salasnich}%
  \BibitemOpen
  \bibfield  {author} {\bibinfo {author} {\bibfnamefont {L.}~\bibnamefont
  {Salasnich}}, \bibinfo {author} {\bibfnamefont {A.}~\bibnamefont {Parola}},\
  and\ \bibinfo {author} {\bibfnamefont {L.}~\bibnamefont {Reatto}},\ }\href
  {https://doi.org/10.1103/PhysRevLett.91.080405} {\bibfield  {journal}
  {\bibinfo  {journal} {Phys. Rev. Lett.}\ }\textbf {\bibinfo {volume} {91}},\
  \bibinfo {pages} {080405} (\bibinfo {year} {2003})}\BibitemShut {NoStop}%
\bibitem [{\citenamefont {Carr}\ and\ \citenamefont {Brand}(2004)}]{Carr2004}%
  \BibitemOpen
  \bibfield  {author} {\bibinfo {author} {\bibfnamefont {L.~D.}\ \bibnamefont
  {Carr}}\ and\ \bibinfo {author} {\bibfnamefont {J.}~\bibnamefont {Brand}},\
  }\href {https://doi.org/10.1103/PhysRevLett.92.040401} {\bibfield  {journal}
  {\bibinfo  {journal} {Phys. Rev. Lett.}\ }\textbf {\bibinfo {volume} {92}},\
  \bibinfo {pages} {040401} (\bibinfo {year} {2004})}\BibitemShut {NoStop}%
\bibitem [{\citenamefont {Glorieux}\ \emph {et~al.}(2025)\citenamefont
  {Glorieux}, \citenamefont {Piekarski}, \citenamefont {Schibler},
  \citenamefont {Aladjidi},\ and\ \citenamefont
  {Baker-Rasooli}}]{glorieux2025}%
  \BibitemOpen
  \bibfield  {author} {\bibinfo {author} {\bibfnamefont {Q.}~\bibnamefont
  {Glorieux}}, \bibinfo {author} {\bibfnamefont {C.}~\bibnamefont {Piekarski}},
  \bibinfo {author} {\bibfnamefont {Q.}~\bibnamefont {Schibler}}, \bibinfo
  {author} {\bibfnamefont {T.}~\bibnamefont {Aladjidi}},\ and\ \bibinfo
  {author} {\bibfnamefont {M.}~\bibnamefont {Baker-Rasooli}},\ }\href
  {https://doi.org/10.1016/bs.aamop.2025.04.002} {\bibfield  {journal}
  {\bibinfo  {journal} {Advances In Atomic, Molecular, and Optical Physics}\
  }\textbf {\bibinfo {volume} {74}},\ \bibinfo {pages} {157} (\bibinfo {year}
  {2025})}\BibitemShut {NoStop}%
\bibitem [{\citenamefont {Carusotto}(2014)}]{Carusotto2014}%
  \BibitemOpen
  \bibfield  {author} {\bibinfo {author} {\bibfnamefont {I.}~\bibnamefont
  {Carusotto}},\ }\href {https://doi.org/10.1098/rspa.2014.0320} {\bibfield
  {journal} {\bibinfo  {journal} {Proceedings. Mathematical, Physical, and
  Engineering Sciences/The Royal Society}\ }\textbf {\bibinfo {volume} {470}},\
  \bibinfo {pages} {20140320} (\bibinfo {year} {2014})}\BibitemShut {NoStop}%
\bibitem [{\citenamefont {Larr\'e}\ and\ \citenamefont
  {Carusotto}(2015)}]{Larre2015}%
  \BibitemOpen
  \bibfield  {author} {\bibinfo {author} {\bibfnamefont {P.-E.}\ \bibnamefont
  {Larr\'e}}\ and\ \bibinfo {author} {\bibfnamefont {I.}~\bibnamefont
  {Carusotto}},\ }\href {https://doi.org/10.1103/PhysRevA.92.043802} {\bibfield
   {journal} {\bibinfo  {journal} {Phys. Rev. A}\ }\textbf {\bibinfo {volume}
  {92}},\ \bibinfo {pages} {043802} (\bibinfo {year} {2015})}\BibitemShut
  {NoStop}%
\bibitem [{SM()}]{SM}%
  \BibitemOpen
  \href@noop {} {\bibinfo {title} {{See Supplemental Material and references
  therein at [URL will be inserted by publisher] for details.}}}\BibitemShut
  {Stop}%
\bibitem [{\citenamefont {Reddy}\ \emph {et~al.}(2015)\citenamefont {Reddy},
  \citenamefont {Permangatt}, \citenamefont {Prabhakar}, \citenamefont {Anwar},
  \citenamefont {Banerji},\ and\ \citenamefont {Singh}}]{Reddy2015}%
  \BibitemOpen
  \bibfield  {author} {\bibinfo {author} {\bibfnamefont {S.~G.}\ \bibnamefont
  {Reddy}}, \bibinfo {author} {\bibfnamefont {C.}~\bibnamefont {Permangatt}},
  \bibinfo {author} {\bibfnamefont {S.}~\bibnamefont {Prabhakar}}, \bibinfo
  {author} {\bibfnamefont {A.}~\bibnamefont {Anwar}}, \bibinfo {author}
  {\bibfnamefont {J.}~\bibnamefont {Banerji}},\ and\ \bibinfo {author}
  {\bibfnamefont {R.}~\bibnamefont {Singh}},\ }\href
  {https://doi.org/https://doi.org/10.1364/AO.56.003556} {\bibfield  {journal}
  {\bibinfo  {journal} {Applied optics}\ }\textbf {\bibinfo {volume} {54}},\
  \bibinfo {pages} {6690} (\bibinfo {year} {2015})}\BibitemShut {NoStop}%
\bibitem [{\citenamefont {Morris}\ \emph {et~al.}(2025)\citenamefont {Morris},
  \citenamefont {Ho}, \citenamefont {Fischer}, \citenamefont {Etrych},
  \citenamefont {Martirosyan}, \citenamefont {Hadzibabic},\ and\ \citenamefont
  {Eigen}}]{Morris2024}%
  \BibitemOpen
  \bibfield  {author} {\bibinfo {author} {\bibfnamefont {S.~J.}\ \bibnamefont
  {Morris}}, \bibinfo {author} {\bibfnamefont {C.~J.}\ \bibnamefont {Ho}},
  \bibinfo {author} {\bibfnamefont {S.~M.}\ \bibnamefont {Fischer}}, \bibinfo
  {author} {\bibfnamefont {J.~c.~v.}\ \bibnamefont {Etrych}}, \bibinfo {author}
  {\bibfnamefont {G.}~\bibnamefont {Martirosyan}}, \bibinfo {author}
  {\bibfnamefont {Z.}~\bibnamefont {Hadzibabic}},\ and\ \bibinfo {author}
  {\bibfnamefont {C.}~\bibnamefont {Eigen}},\ }\href
  {https://doi.org/10.1103/PhysRevA.111.L041301} {\bibfield  {journal}
  {\bibinfo  {journal} {Phys. Rev. A}\ }\textbf {\bibinfo {volume} {111}},\
  \bibinfo {pages} {L041301} (\bibinfo {year} {2025})}\BibitemShut {NoStop}%
\end{thebibliography}%


\begin{thebibliography}{12}%
\makeatletter
\providecommand \@ifxundefined [1]{%
 \@ifx{#1\undefined}
}%
\providecommand \@ifnum [1]{%
 \ifnum #1\expandafter \@firstoftwo
 \else \expandafter \@secondoftwo
 \fi
}%
\providecommand \@ifx [1]{%
 \ifx #1\expandafter \@firstoftwo
 \else \expandafter \@secondoftwo
 \fi
}%
\providecommand \natexlab [1]{#1}%
\providecommand \enquote  [1]{``#1''}%
\providecommand \bibnamefont  [1]{#1}%
\providecommand \bibfnamefont [1]{#1}%
\providecommand \citenamefont [1]{#1}%
\providecommand \href@noop [0]{\@secondoftwo}%
\providecommand \href [0]{\begingroup \@sanitize@url \@href}%
\providecommand \@href[1]{\@@startlink{#1}\@@href}%
\providecommand \@@href[1]{\endgroup#1\@@endlink}%
\providecommand \@sanitize@url [0]{\catcode `\\12\catcode `\$12\catcode
  `\&12\catcode `\#12\catcode `\^12\catcode `\_12\catcode `\%12\relax}%
\providecommand \@@startlink[1]{}%
\providecommand \@@endlink[0]{}%
\providecommand \url  [0]{\begingroup\@sanitize@url \@url }%
\providecommand \@url [1]{\endgroup\@href {#1}{\urlprefix }}%
\providecommand \urlprefix  [0]{URL }%
\providecommand \Eprint [0]{\href }%
\providecommand \doibase [0]{https://doi.org/}%
\providecommand \selectlanguage [0]{\@gobble}%
\providecommand \bibinfo  [0]{\@secondoftwo}%
\providecommand \bibfield  [0]{\@secondoftwo}%
\providecommand \translation [1]{[#1]}%
\providecommand \BibitemOpen [0]{}%
\providecommand \bibitemStop [0]{}%
\providecommand \bibitemNoStop [0]{.\EOS\space}%
\providecommand \EOS [0]{\spacefactor3000\relax}%
\providecommand \BibitemShut  [1]{\csname bibitem#1\endcsname}%
\let\auto@bib@innerbib\@empty
\bibitem [{\citenamefont {Carusotto}\ and\ \citenamefont
  {Ciuti}(2013)}]{Carusotto2013}%
  \BibitemOpen
  \bibfield  {author} {\bibinfo {author} {\bibfnamefont {I.}~\bibnamefont
  {Carusotto}}\ and\ \bibinfo {author} {\bibfnamefont {C.}~\bibnamefont
  {Ciuti}},\ }\href {https://doi.org/10.1103/RevModPhys.85.299} {\bibfield
  {journal} {\bibinfo  {journal} {Rev. Mod. Phys.}\ }\textbf {\bibinfo {volume}
  {85}},\ \bibinfo {pages} {299} (\bibinfo {year} {2013})}\BibitemShut
  {NoStop}%
\bibitem [{\citenamefont {Glorieux}\ \emph {et~al.}(2023)\citenamefont
  {Glorieux}, \citenamefont {Aladjidi}, \citenamefont {Lett},\ and\
  \citenamefont {Kaiser}}]{Glorieux2023}%
  \BibitemOpen
  \bibfield  {author} {\bibinfo {author} {\bibfnamefont {Q.}~\bibnamefont
  {Glorieux}}, \bibinfo {author} {\bibfnamefont {T.}~\bibnamefont {Aladjidi}},
  \bibinfo {author} {\bibfnamefont {P.~D.}\ \bibnamefont {Lett}},\ and\
  \bibinfo {author} {\bibfnamefont {R.}~\bibnamefont {Kaiser}},\ }\href@noop {}
  {\bibfield  {journal} {\bibinfo  {journal} {New Journal of Physics}\ }\textbf
  {\bibinfo {volume} {25}},\ \bibinfo {pages} {051201} (\bibinfo {year}
  {2023})}\BibitemShut {NoStop}%
\bibitem [{\citenamefont {Carusotto}(2014)}]{carusotto2014superfluid}%
  \BibitemOpen
  \bibfield  {author} {\bibinfo {author} {\bibfnamefont {I.}~\bibnamefont
  {Carusotto}},\ }\href@noop {} {\bibfield  {journal} {\bibinfo  {journal}
  {Proceedings of the Royal Society A: Mathematical, Physical and Engineering
  Sciences}\ }\textbf {\bibinfo {volume} {470}},\ \bibinfo {pages} {20140320}
  (\bibinfo {year} {2014})}\BibitemShut {NoStop}%
\bibitem [{\citenamefont {Dennis}\ \emph {et~al.}(2013)\citenamefont {Dennis},
  \citenamefont {Hope},\ and\ \citenamefont {Johnsson}}]{DENNIS2013201}%
  \BibitemOpen
  \bibfield  {author} {\bibinfo {author} {\bibfnamefont {G.~R.}\ \bibnamefont
  {Dennis}}, \bibinfo {author} {\bibfnamefont {J.~J.}\ \bibnamefont {Hope}},\
  and\ \bibinfo {author} {\bibfnamefont {M.~T.}\ \bibnamefont {Johnsson}},\
  }\href {https://doi.org/https://doi.org/10.1016/j.cpc.2012.08.016} {\bibfield
   {journal} {\bibinfo  {journal} {Computer Physics Communications}\ }\textbf
  {\bibinfo {volume} {184}},\ \bibinfo {pages} {201} (\bibinfo {year}
  {2013})}\BibitemShut {NoStop}%
\bibitem [{\citenamefont {Pitaevskii}\ and\ \citenamefont
  {Stringari}(2016)}]{Pitaevskii2016}%
  \BibitemOpen
  \bibfield  {author} {\bibinfo {author} {\bibfnamefont {L.}~\bibnamefont
  {Pitaevskii}}\ and\ \bibinfo {author} {\bibfnamefont {S.}~\bibnamefont
  {Stringari}},\ }\href@noop {} {\emph {\bibinfo {title} {Bose-Einstein
  condensation and superfluidity}}},\ Vol.\ \bibinfo {volume} {164}\ (\bibinfo
  {publisher} {Oxford University Press},\ \bibinfo {year} {2016})\BibitemShut
  {NoStop}%
\bibitem [{\citenamefont {Taylor}\ and\ \citenamefont
  {Zaremba}(2003)}]{PhysRevA.68.053611}%
  \BibitemOpen
  \bibfield  {author} {\bibinfo {author} {\bibfnamefont {E.}~\bibnamefont
  {Taylor}}\ and\ \bibinfo {author} {\bibfnamefont {E.}~\bibnamefont
  {Zaremba}},\ }\href {https://doi.org/10.1103/PhysRevA.68.053611} {\bibfield
  {journal} {\bibinfo  {journal} {Phys. Rev. A}\ }\textbf {\bibinfo {volume}
  {68}},\ \bibinfo {pages} {053611} (\bibinfo {year} {2003})}\BibitemShut
  {NoStop}%
\bibitem [{\citenamefont {Kr{\"a}mer}\ \emph {et~al.}(2005)\citenamefont
  {Kr{\"a}mer}, \citenamefont {Menotti},\ and\ \citenamefont
  {Modugno}}]{Kramer2005}%
  \BibitemOpen
  \bibfield  {author} {\bibinfo {author} {\bibfnamefont {M.}~\bibnamefont
  {Kr{\"a}mer}}, \bibinfo {author} {\bibfnamefont {C.}~\bibnamefont
  {Menotti}},\ and\ \bibinfo {author} {\bibfnamefont {M.}~\bibnamefont
  {Modugno}},\ }\href {https://doi.org/10.1007/s10909-005-2294-z} {\bibfield
  {journal} {\bibinfo  {journal} {Journal of Low Temperature Physics}\ }\textbf
  {\bibinfo {volume} {138}},\ \bibinfo {pages} {729} (\bibinfo {year}
  {2005})}\BibitemShut {NoStop}%
\bibitem [{\citenamefont {Nguyen}\ \emph {et~al.}(2017)\citenamefont {Nguyen},
  \citenamefont {Luo},\ and\ \citenamefont {Hulet}}]{2017Nguyen}%
  \BibitemOpen
  \bibfield  {author} {\bibinfo {author} {\bibfnamefont {J.~H.~V.}\
  \bibnamefont {Nguyen}}, \bibinfo {author} {\bibfnamefont {D.}~\bibnamefont
  {Luo}},\ and\ \bibinfo {author} {\bibfnamefont {R.~G.}\ \bibnamefont
  {Hulet}},\ }\href {https://doi.org/10.1126/science.aal3220} {\bibfield
  {journal} {\bibinfo  {journal} {Science}\ }\textbf {\bibinfo {volume}
  {356}},\ \bibinfo {pages} {422} (\bibinfo {year} {2017})}\BibitemShut
  {NoStop}%
\bibitem [{\citenamefont {Salasnich}\ \emph {et~al.}(2003)\citenamefont
  {Salasnich}, \citenamefont {Parola},\ and\ \citenamefont
  {Reatto}}]{2003Salasnich}%
  \BibitemOpen
  \bibfield  {author} {\bibinfo {author} {\bibfnamefont {L.}~\bibnamefont
  {Salasnich}}, \bibinfo {author} {\bibfnamefont {A.}~\bibnamefont {Parola}},\
  and\ \bibinfo {author} {\bibfnamefont {L.}~\bibnamefont {Reatto}},\ }\href
  {https://doi.org/10.1103/PhysRevLett.91.080405} {\bibfield  {journal}
  {\bibinfo  {journal} {Phys. Rev. Lett.}\ }\textbf {\bibinfo {volume} {91}},\
  \bibinfo {pages} {080405} (\bibinfo {year} {2003})}\BibitemShut {NoStop}%
\bibitem [{\citenamefont {Fibich}\ and\ \citenamefont
  {Gaeta}(2000)}]{Fibich2000}%
  \BibitemOpen
  \bibfield  {author} {\bibinfo {author} {\bibfnamefont {G.}~\bibnamefont
  {Fibich}}\ and\ \bibinfo {author} {\bibfnamefont {A.~L.}\ \bibnamefont
  {Gaeta}},\ }\href@noop {} {\bibfield  {journal} {\bibinfo  {journal} {Optics
  letters}\ }\textbf {\bibinfo {volume} {25}},\ \bibinfo {pages} {335}
  (\bibinfo {year} {2000})}\BibitemShut {NoStop}%
\bibitem [{\citenamefont {Vuong}\ \emph {et~al.}(2006)\citenamefont {Vuong},
  \citenamefont {Grow}, \citenamefont {Ishaaya}, \citenamefont {Gaeta},
  \citenamefont {'t~Hooft}, \citenamefont {Eliel},\ and\ \citenamefont
  {Fibich}}]{vuong2006}%
  \BibitemOpen
  \bibfield  {author} {\bibinfo {author} {\bibfnamefont {L.~T.}\ \bibnamefont
  {Vuong}}, \bibinfo {author} {\bibfnamefont {T.~D.}\ \bibnamefont {Grow}},
  \bibinfo {author} {\bibfnamefont {A.}~\bibnamefont {Ishaaya}}, \bibinfo
  {author} {\bibfnamefont {A.~L.}\ \bibnamefont {Gaeta}}, \bibinfo {author}
  {\bibfnamefont {G.~W.}\ \bibnamefont {'t~Hooft}}, \bibinfo {author}
  {\bibfnamefont {E.~R.}\ \bibnamefont {Eliel}},\ and\ \bibinfo {author}
  {\bibfnamefont {G.}~\bibnamefont {Fibich}},\ }\href
  {https://doi.org/10.1103/PhysRevLett.96.133901} {\bibfield  {journal}
  {\bibinfo  {journal} {Phys. Rev. Lett.}\ }\textbf {\bibinfo {volume} {96}},\
  \bibinfo {pages} {133901} (\bibinfo {year} {2006})}\BibitemShut {NoStop}%
\bibitem [{\citenamefont {Kruglov}\ \emph {et~al.}(1992)\citenamefont
  {Kruglov}, \citenamefont {Logvin},\ and\ \citenamefont
  {Volkov}}]{kruglov1992}%
  \BibitemOpen
  \bibfield  {author} {\bibinfo {author} {\bibfnamefont {V.}~\bibnamefont
  {Kruglov}}, \bibinfo {author} {\bibfnamefont {Y.~A.}\ \bibnamefont
  {Logvin}},\ and\ \bibinfo {author} {\bibfnamefont {V.}~\bibnamefont
  {Volkov}},\ }\href@noop {} {\bibfield  {journal} {\bibinfo  {journal}
  {Journal of Modern Optics}\ }\textbf {\bibinfo {volume} {39}},\ \bibinfo
  {pages} {2277} (\bibinfo {year} {1992})}\BibitemShut {NoStop}%
\end{thebibliography}%


\begin{thebibliography}{0}%
\makeatletter
\providecommand \@ifxundefined [1]{%
 \@ifx{#1\undefined}
}%
\providecommand \@ifnum [1]{%
 \ifnum #1\expandafter \@firstoftwo
 \else \expandafter \@secondoftwo
 \fi
}%
\providecommand \@ifx [1]{%
 \ifx #1\expandafter \@firstoftwo
 \else \expandafter \@secondoftwo
 \fi
}%
\providecommand \natexlab [1]{#1}%
\providecommand \enquote  [1]{``#1''}%
\providecommand \bibnamefont  [1]{#1}%
\providecommand \bibfnamefont [1]{#1}%
\providecommand \citenamefont [1]{#1}%
\providecommand \href@noop [0]{\@secondoftwo}%
\providecommand \href [0]{\begingroup \@sanitize@url \@href}%
\providecommand \@href[1]{\@@startlink{#1}\@@href}%
\providecommand \@@href[1]{\endgroup#1\@@endlink}%
\providecommand \@sanitize@url [0]{\catcode `\\12\catcode `\$12\catcode
  `\&12\catcode `\#12\catcode `\^12\catcode `\_12\catcode `\%12\relax}%
\providecommand \@@startlink[1]{}%
\providecommand \@@endlink[0]{}%
\providecommand \url  [0]{\begingroup\@sanitize@url \@url }%
\providecommand \@url [1]{\endgroup\@href {#1}{\urlprefix }}%
\providecommand \urlprefix  [0]{URL }%
\providecommand \Eprint [0]{\href }%
\providecommand \doibase [0]{https://doi.org/}%
\providecommand \selectlanguage [0]{\@gobble}%
\providecommand \bibinfo  [0]{\@secondoftwo}%
\providecommand \bibfield  [0]{\@secondoftwo}%
\providecommand \translation [1]{[#1]}%
\providecommand \BibitemOpen [0]{}%
\providecommand \bibitemStop [0]{}%
\providecommand \bibitemNoStop [0]{.\EOS\space}%
\providecommand \EOS [0]{\spacefactor3000\relax}%
\providecommand \BibitemShut  [1]{\csname bibitem#1\endcsname}%
\let\auto@bib@innerbib\@empty
\end{thebibliography}%
\end{document}